\documentclass[10pt,conference]{IEEEtran}
\usepackage{cite}
\usepackage{amsmath,amssymb,amsfonts}
\usepackage{algorithmic}
\usepackage{graphicx}
\usepackage{textcomp}
\usepackage{xcolor}
\usepackage[hyphens]{url}
\usepackage{fancyhdr}
\usepackage{tikz}
\usepackage{xcolor}
\usepackage{multirow} 
\usepackage{tabularx}
\usepackage{booktabs}
\newcommand*\circled[1]{\tikz[baseline=(char.base)]{
            \node[shape=circle,fill,inner sep=0.3pt] (char) {\textcolor{white}{#1}};}}
            
\usepackage{soul}

\newcommand{\hpcayear}{2025}

\title{\huge Panacea: Novel DNN Accelerator using Accuracy-Preserving Asymmetric Quantization and Energy-Saving Bit-Slice Sparsity\!}

\newcommand\hpcaauthors{
    Dongyun~Kam$^{\dagger}$,
    Myeongji~Yun$^{\dagger}$,
    Sunwoo~Yoo$^{\dagger}$,
    Seungwoo~Hong$^{\dagger}$,
    Zhengya~Zhang$^{\ddagger}$,
    and~Youngjoo~Lee$^{\dagger}$}
\newcommand\hpcaaffiliation{Pohang University of Science and Technology (POSTECH)$^{\dagger}$, University of Michigan$^{\ddagger}$}
\newcommand\hpcaemail{\{rkaehddbs, mjyun01, swyoo23, seungwoohong, youngjoo.lee\}@postech.ac.kr$^{\dagger}$, zhengya@umich.edu$^{\ddagger}$}

\author{
  \ifdefined\hpcacameraready
    \IEEEauthorblockN{\hpcaauthors{}}
      \IEEEauthorblockA{
        \hpcaaffiliation{} \\
        \hpcaemail{}
      }
  \else
      \IEEEauthorblockN{\hpcaauthors{}}
      \IEEEauthorblockA{
        \hpcaaffiliation{} \\
        \hpcaemail{}
      }
  \fi 
}

\fancypagestyle{camerareadyfirstpage}{%
  \fancyhead{}
  
  \fancyhead[C]{
    \ifdefined\aeopen
    \parbox[][12mm][t]{13.5cm}{\hpcayear{} IEEE International Symposium on High-Performance Computer Architecture (HPCA)}    
    \else
      \ifdefined\aereviewed
      \parbox[][12mm][t]{13.5cm}{\hpcayear{} IEEE International Symposium on High-Performance Computer Architecture (HPCA)}
      \else
      \ifdefined\aereproduced
      \parbox[][12mm][t]{13.5cm}{\hpcayear{} IEEE International Symposium on High-Performance Computer Architecture (HPCA)}
      \else
      \parbox[][0mm][t]{13.5cm}{\hpcayear{} IEEE International Symposium on High-Performance Computer Architecture (HPCA)}
    \fi 
    \fi 
    \fi 
    \ifdefined\aeopen 
      \includegraphics[width=12mm,height=12mm]{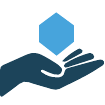}
    \fi 
    \ifdefined\aereviewed
      \includegraphics[width=12mm,height=12mm]{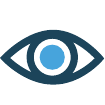}
    \fi 
    \ifdefined\aereproduced
      \includegraphics[width=12mm,height=12mm]{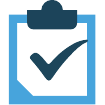}
    \fi
  }
  \fancyfoot[C]{}
}
\begin{document}
\maketitle

\ifdefined\hpcacameraready 
  \thispagestyle{camerareadyfirstpage}
  \pagestyle{empty}
\else
  \thispagestyle{plain}
  \pagestyle{plain}
\fi

\newcommand{\hpcaheight}{0mm}
\ifdefined\eaopen
\renewcommand{\hpcaheight}{12mm}
\fi


\begin{abstract}
Low bit-precisions and their bit-slice sparsity have recently been studied to accelerate general matrix-multiplications (GEMM) during large-scale deep neural network (DNN) inferences.
While the conventional symmetric quantization facilitates low-resolution processing with bit-slice sparsity for both weight and activation, its accuracy loss caused by the activation's asymmetric distributions cannot be acceptable, especially for large-scale DNNs. 
In efforts to mitigate this accuracy loss, recent studies have actively utilized asymmetric quantization for activations without requiring additional operations. 
However, the cutting-edge asymmetric quantization produces numerous nonzero slices that cannot be compressed and skipped by recent bit-slice GEMM accelerators, naturally consuming more processing energy to handle the quantized DNN models.

To simultaneously achieve high accuracy and hardware efficiency for large-scale DNN inferences, this paper proposes an Asymmetrically-Quantized bit-Slice GEMM (AQS-GEMM) for the first time.
In contrast to the previous bit-slice computing, which only skips operations of zero slices, the AQS-GEMM compresses frequent nonzero slices, generated by asymmetric quantization, and skips their operations.
To increase the slice-level sparsity of activations, we also introduce two algorithm-hardware co-optimization methods: a zero-point manipulation and a distribution-based bit-slicing.
To support the proposed AQS-GEMM and optimizations at the hardware-level, we newly introduce a DNN accelerator, \textit{Panacea}, which efficiently handles sparse/dense workloads of the tiled AQS-GEMM to increase data reuse and utilization.
\textit{Panacea} supports a specialized dataflow and run-length encoding to maximize data reuse and minimize external memory accesses, significantly improving its hardware efficiency.
Numerous benchmark evaluations show that \textit{Panacea} outperforms existing DNN accelerators, e.g., 1.97$\times$ and 3.26$\times$ higher energy efficiency, and 1.88$\times$ and 2.41$\times$ higher throughput than the recent bit-slice accelerator \textit{Sibia} and the SIMD design, respectively, on OPT-2.7B, while providing better algorithm performance with asymmetric quantization.
\end{abstract}

\section{Introduction}

Large-scale deep neural networks (DNNs) have been employed as powerful solutions for a variety of practical applications, including image classification \cite{he2016deep,dosovitskiy2020image,touvron2021training}, natural language processing (NLP) \cite{wang2015learning, devlin2018bert, radford2018improving, radford2019language,floridi2020gpt, touvron2023llama}, AR/VR \cite{zhou2019edge}, and robotics \cite{park2020deep}.
However, their extensive computations and external memory accesses (EMA) result in poor throughput and increased energy consumption during inferences \cite{sastry2024computing, kaplan2020scaling, cao2020towards, kim2023full}, leading to more significant challenges for a large-scale DNN's deployment on resource-constrained edge devices \cite{wang2022survey, hussain2022design}.

\begin{figure}[t]
    \centering
    \includegraphics{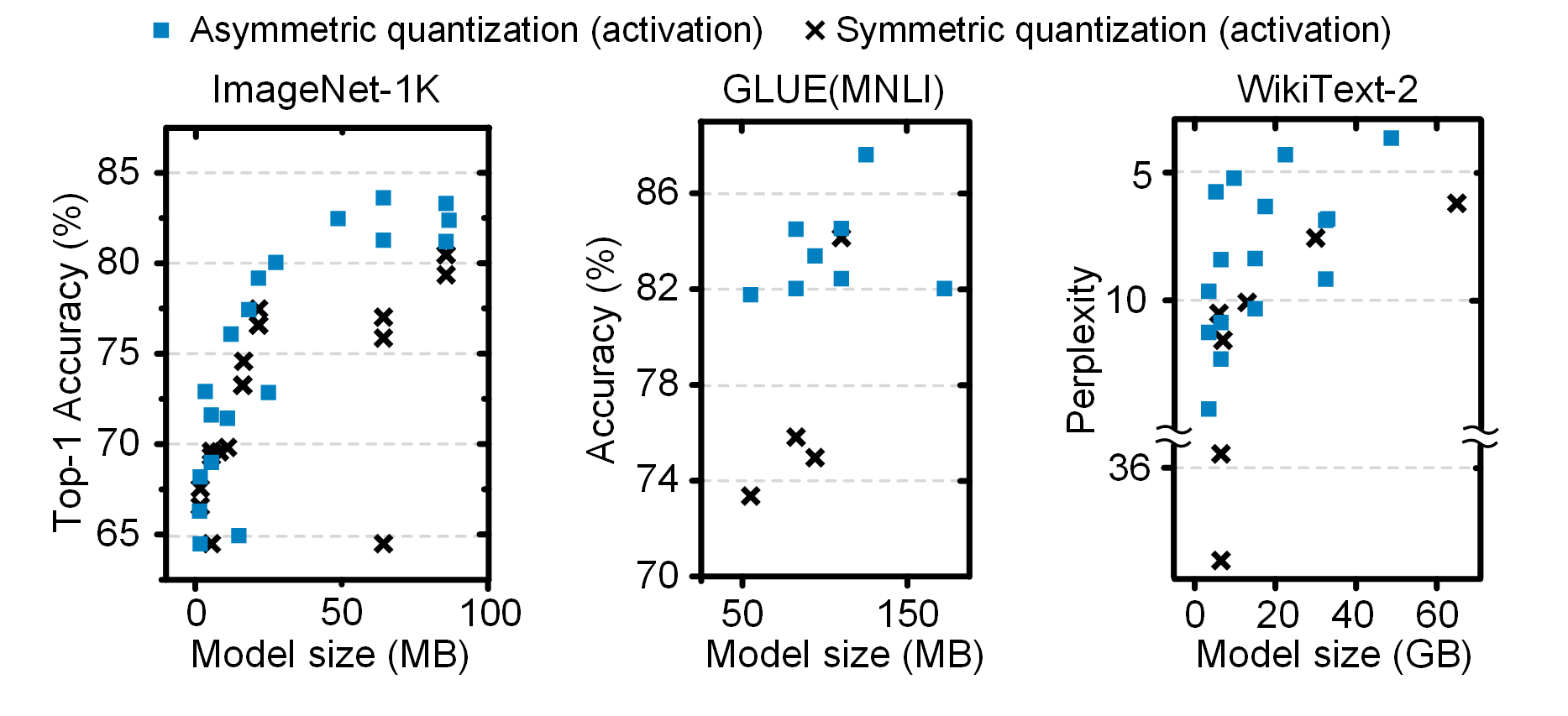}
    \vspace*{-6mm}
    \caption{Accuracy comparison on recent works utilizing symmetric \cite{wu2020easyquant, liu2021post, li2021brecq, banner2018aciq, xiao2023smoothquant} and asymmetric quantization \cite{lin2021fq, liu2023pd, lee2023flexround, cai2020zeroq, li2023repq, wei2022outlier, wei2023outlier, nagel2021white, liu2023qllm, shao2023omniquant} for activations in large-scale DNNs.}
    \label{introduction}
    \vspace*{-3mm}
\end{figure}

To address these challenges, recent works \cite{shi2024bitwave, zhang2016cambricon, zhu2019sparse, han2015learning, fang2022algorithm, min2023energy, byun2023sparsity, frantar2023sparsegpt, 
kim2021bert, jacob2018quantization, guo2023olive, dai2021vs, zadeh2020gobo, frantar2022optq, yuan2023rptq, guo2022ant} have developed energy-efficient DNN inference accelerators by employing algorithm-hardware co-optimization techniques such as weight pruning \cite{shi2024bitwave, zhang2016cambricon, zhu2019sparse, han2015learning, fang2022algorithm, min2023energy, byun2023sparsity, frantar2023sparsegpt} and quantization \cite{ zhou2016dorefa, kim2021bert, jacob2018quantization, guo2023olive, zadeh2020gobo, dai2021vs, frantar2022optq, yuan2023rptq, guo2022ant }.
Among these techniques, low bit-precision post-training quantization (PTQ) \cite{guo2023olive, zadeh2020gobo, dai2021vs, frantar2022optq, yuan2023rptq} stands out by significantly lowering the complexity of operations and memory accesses, even without necessitating costly fine-tuning with large datasets.
This results from enabling integer general matrix-multiplications (GEMMs) by quantizing weights and activations.
However, PTQ-based methods, e.g., symmetric and asymmetric quantization, still have challenges for large-scale DNN inferences on resource-constrained devices, as summarized below.

\noindent{\textbf{Accuracy degradation of symmetric quantization:}}
While symmetric quantization generally enables low bit-precisions, it can cause significant accuracy loss during large-scale DNN inferences, such as transformer models, which contain layers producing asymmetric distributions or long-tail distributions in input activation matrices \cite{liu2023noisyquant, yuan2022ptq4vit}.
Recent studies at the algorithm-level \cite{wu2020easyquant, liu2021post, li2021brecq, banner2018aciq, xiao2023smoothquant, lin2021fq, liu2023pd, lee2023flexround, cai2020zeroq, li2023repq, wei2022outlier, wei2023outlier, liu2023qllm, shao2023omniquant, nagel2021white } have addressed the accuracy degradation issue by applying symmetric quantization to weights and asymmetric quantization to activations, as illustrated in Fig. \ref{introduction}.
Consequently, to capitalize on the large-scale model's high accuracy fully, it is crucial to utilize low-bit asymmetric quantization for input activations.

\begin{figure}
    \centering
    \includegraphics{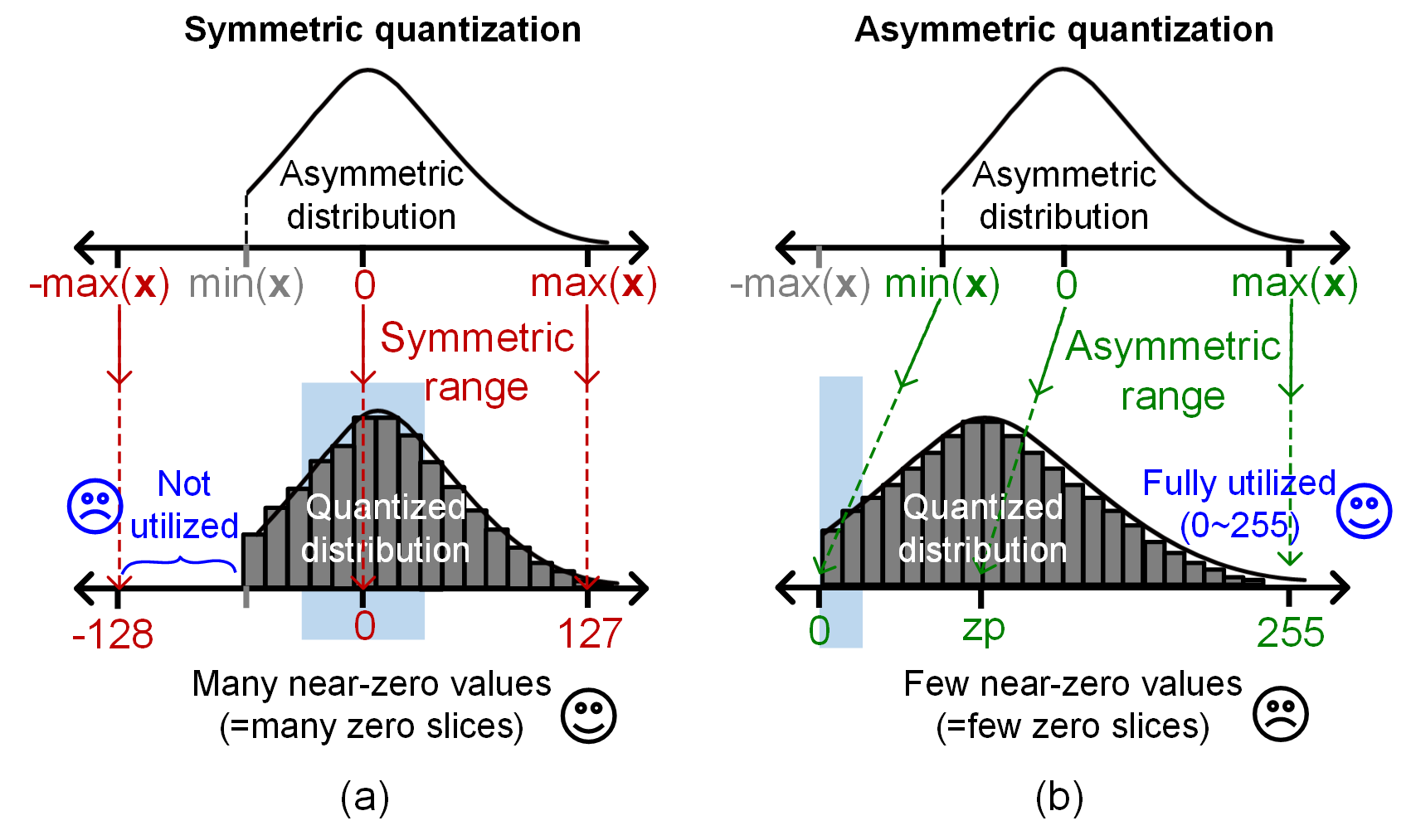}
    \vspace*{-3mm}
    \caption{Examples of uniform quantization methods: 8-bit (a) symmetric and (b) asymmetric approaches.}
    \vspace*{-3mm}
    \label{distribution}
\end{figure}

\noindent{\textbf{HW optimization challenge of integer GEMMs with asymmetric quantization:}}
However, integrating integer GEMMs with asymmetric quantization makes it difficult to utilize recent optimization techniques, such as sparsity, for energy-efficient DNN inferences \cite{lin2024qserve, nagel2021white}.
While integer GEMMs with symmetric quantization \cite{im2024sibia, shomron2020non, han2023hnpu, im2024lutein}, have actively supported sparsity even for dense DNN models, which have high bit-slice sparsity ($>$90\%) in high-order slices after segmenting near-zero values into multiple bit-slices, integer GEMMs with asymmetric quantization cannot exploit the slice sparsity due to few zero slices.
Unlike symmetric quantization centering values around zero, as shown in Fig. \ref{distribution}, asymmetric quantization produces quantized distributions that are not centered around zero, generating many nonzero slices that cannot be directly compressed and skipped.
Therefore, to simultaneously enhance accuracy and hardware efficiency, it is essential that a new GEMM operation and its hardware architecture efficiently address the numerous multiply-and-accumulate (MAC) operations and memory accesses associated with nonzero slices.

This paper proposes the first asymmetrically-quantized bit-slice DNN accelerator, \textit{Panacea}, successfully achieving high accuracy and hardware efficiency for resource-limited devices.
The contributions of \textit{Panacea} are summarized as follows:

\begin{itemize}
    \item 
    The \textbf{A}symmetrically-\textbf{Q}uantized bit-\textbf{S}lice \textbf{GEMM} (AQS-GEMM) is newly developed to compress high-order nonzero slices, which are frequently observed in asymmetric quantization, and skip MAC operations of compressed slices.
    To get exact GEMM results, the AQS-GEMM efficiently computes a compensation term for skipping compressed slices, while reusing loaded data and minimizing additional memory accesses.
    It finally reduces the number of MAC operations by 61\% compared to the dense GEMM, enabling bit-slice sparsity.
    \item
    To increase high-order (HO) slice sparsity in activations, \textit{Panacea} utilizes two algorithm-hardware co-optimization methods: \textbf{Z}ero-\textbf{P}oint \textbf{M}anipulation (ZPM) and \textbf{D}istribution-based \textbf{B}it-\textbf{S}licing (DBS).
    They are activated during the PTQ calibration (algorithm-level) and exploit higher sparsity during inferences (hardware-level).
    The ZPM properly adjusts the zero-point value of activation to put more quantized values into a slice skip range, thereby increasing the HO slice sparsity by at most 33\%p.
    To further increase the sparsity, the DBS categorizes quantized distributions into three types during the calibration.
    Based on the identified types, the DBS applies different bit-slicing rules to different layers during the inference step by using low-cost shifting units, increasing the HO slice sparsity by at most 56\%p and overall energy efficiency by 10.1\% for DeiT-base.
    
    \item 
    \textit{Panacea} efficiently computes the AQS-GEMM with the run-length encoded bit-slices, reducing the number of MAC operations and EMA.
    Its processing elements maximize data reuse by dedicating sparse and dense workload operators, which are activated in parallel, further reducing on-chip memory accesses.
    The processing elements' compensator also reuses the loaded weight slices to compensate for the skipped MAC operations, producing exact GEMM results with negligible overhead.
    To address the low utilization problem of resources at high slice sparsity, \textit{Panacea} adopts a double-tile processing flow that allocates additional computation workload into free operators in each processing element.   
\end{itemize}

We evaluate the hardware-level performance of \textit{Panacea} on different design parameters, such as the number of operators, and different slice sparsities of weights and activations.
In addition, to properly determine \textit{Panacea}'s design options, we analyze the slice sparsity for large-scale DNN benchmark models.
Our evaluations demonstrate that \textit{Panacea} outperforms recent dense DNN accelerators: the systolic array \cite{ verhelst2017embedded, asgari2020meissa}, the SIMD \cite{keller202395}, and the bit-slice accelerator \textit{Sibia} \cite{im2024sibia}.
Compared to \textit{Sibia}\cite{im2024sibia}, which only supports symmetric quantization, \textit{Panacea} improves energy efficiency by 2.03$\times$, 1.97$\times$, 1.52$\times$, and 1.49$\times$ and throughput by 1.34$\times$, 1.88$\times$, 1.24$\times$, and 1.37$\times$ for GPT-2 \cite{radford2019language}, OPT-2.7B \cite{zhang2022opt}, Llama-3.2-1B \cite{llama3.2}, and ResNet-18 \cite{he2016deep}, respectively, by efficiently utilizing high slice sparsity, while providing better accuracy with asymmetric quantization.
The results also show \textit{Panacea} outperforms \textit{Sibia} for OPT-2.7B \cite{zhang2022opt} using 4-bit weights.

\section{Background \& Motivation}

\subsection{Quantization and integer GEMM}
\noindent\textbf{Uniform Quantization.}
Quantization plays a pivotal role in reducing the computational cost and memory footprint by enabling computations with low bit-precision \cite{kim2021bert, jacob2018quantization, guo2023olive, dai2021vs, zadeh2020gobo,  frantar2022optq, yuan2023rptq, zhou2016dorefa, wu2022xtc, park2018energy, song2020drq, jang2024figna, nahshan2021loss, sze2017efficient, nagel2021white, gholami2022survey}.
There are two uniform quantization techniques: symmetric and asymmetric quantization schemes, which usually use signed and unsigned integers, respectively.
Given an input matrix $\textbf{x}$, two schemes are computed as follows:
\begin{align}
    \textbf{x}_{\text{int}b} & = Q(\textbf{x};s,b) = \text{clip}(\left\lfloor{\frac{\textbf{x}}{s}} \right\rceil;-2^{b-1},2^{b-1}-1), \\
    \textbf{x}_{\text{uint}b} & = Q(\textbf{x};s',zp,b) = \text{clip}(\left\lfloor{\frac{\textbf{x}}{s'}} \right\rceil+zp;0,2^{b}-1),
\end{align}
where quantization parameters are defined as the scale factors $\textit{s} = 2\times \text{max}(|\textbf{x}|)/(2^{b}-1)$, $\textit{s}' = (\text{max}(\textbf{x})-\text{min}(\textbf{x}))/(2^{b}-1)$, the bit-width $\textit{b}$, and the $b$-bit zero-point $\textit{zp} = \text{clip}(\left\lfloor{-\text{min}(\textbf{x})/\textit{s}'}\right\rceil;0,2^b-1)$.
Essentially, in symmetric quantization, the positive and negative ranges of a value are symmetrically mapped to the positive and negative sides of the quantized space \cite{nagel2021white, gholami2022survey}, as shown in Fig. \ref{distribution}(a). 
This often leads to underutilization of part of the quantized space, like in common scenarios of DNN computation where the negative quantized space is not fully utilized. 
On the other hand, asymmetric quantization maps the entire value range to the unsigned quantized space, mapping the zero value into $zp$ that no longer centers the quantized space but instead varies based on the range of the value \cite{nagel2021white, gholami2022survey}, as shown in Fig. \ref{distribution} (b). 
This form of quantization provides the best representation of the value for a given bit-width, accommodating the actual data distribution more flexibly.
Using a unique zero-point value for an input matrix enhances the precision with which distribution characteristics are captured, thus reducing the effects of precision loss \cite{lin2021fq, liu2023pd, lee2023flexround, cai2020zeroq, li2023repq, wei2022outlier, wei2023outlier, nagel2021white, liu2023qllm, shao2023omniquant}.

\noindent\textbf{Post-Training Quantization.}
PTQ quantizes weights and input activations without the costly fine-tuning and any labeled dataset.
This method only necessitates a minimal calibration dataset, typically a subset of the training dataset \cite{hubara2021accurate}.
During the calibration, this subset is fed into a DNN model to calculate each layer's scale factor and zero point \cite{hubara2021accurate}.
Consequently, PTQ offers an efficient and rapid quantization approach, which is especially beneficial for the varied landscape of DNN architectures.
In this paper, we consistently apply PTQ across all weights and input activations. 
\begin{figure}[!t]
    \centering
    \vspace*{-3mm}
    \includegraphics{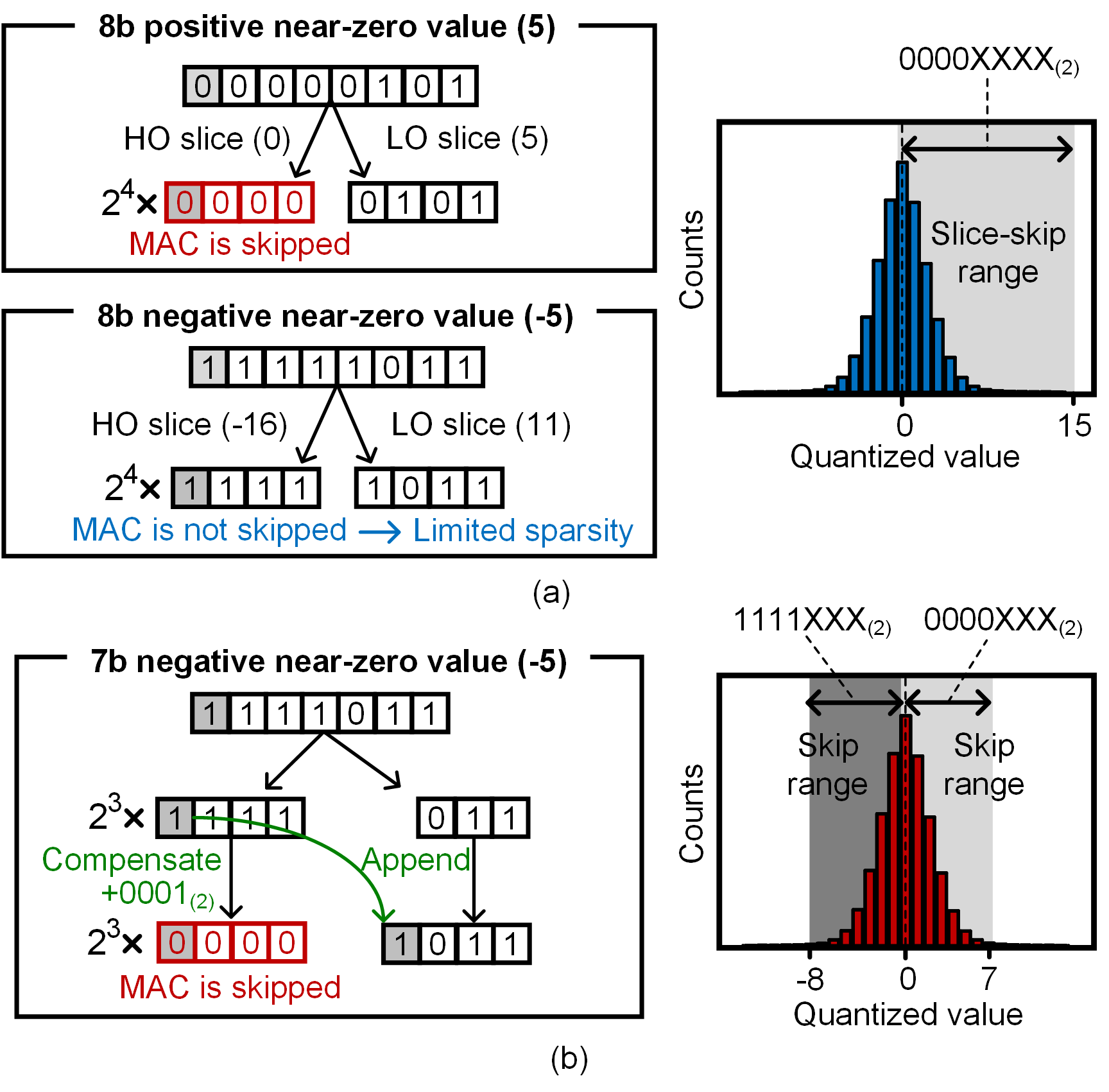}
    \vspace*{-3mm}
    \caption{(a) the straightforward bit-slice representation \cite{shomron2020non}, and (b) the signed bit-slice representation \cite{im2024sibia}.}
    \vspace*{-2mm}
    \label{slicing}
\end{figure}

\noindent\textbf{Integer GEMM with asymmetric quantization.}
Large-scale DNNs often incorporate a lot of computationally intensive GEMMs \cite{sze2017efficient}.
Quantizing weights and activations in all layers makes it possible to employ low-bit integer GEMMs, resulting in significantly enhancing hardware efficiency. 
To minimize accuracy degradation caused by the PTQ without additional computational overhead, one effective strategy is to compute GEMMs with symmetric weight quantization and asymmetric activation quantization \cite{nagel2021white} as detailed in (3).
\begin{align}
    \textbf{W}\textbf{x} + \textbf{b} & \approx s_{\text{W}}(\textbf{W}_{\text{int}})s_{\text{x}}(\textbf{x}_{\text{uint}}-zp_{\text{x}}) + \textbf{b} \notag \\ 
    & = s_{\text{W}} s_{\text{x}} (\textbf{W}_{\text{int}}\textbf{x}_{\text{uint}}-zp_{\text{x}}\textbf{W}_{\text{int}}\textbf{1}^{K \times 1} + \textbf{b}_{\text{int}}) \notag \\
    & = s_{\text{W}} s_{\text{x}} (\textbf{W}_{\text{int}}\textbf{x}_{\text{uint}} + \hat{\textbf{b}}_{\text{int}}),
\end{align}
where $\textbf{W}$ is a $M \times K$ weight, $\textbf{x}$ is a $K \times N$ activation, $\textbf{b}$ is a $M \times 1$ bias, and $\textbf{1}^{K \times 1}$ is a $K \times 1$ vector consisting entirely of ones.
The term $\textbf{W}_{\text{int}}\textbf{x}_{\text{uint}}$ is an integer GEMM that is mainly performed in a DNN accelerator.
The second term $zp_{\text{x}}\textbf{W}_{\text{int}}\textbf{1}^{K \times 1}$ can be pre-computed and added to the bias term because it is independent on $\textbf{x}_{\text{uint}}$ and known in advance.
Thus, computing the integer GEMMs with asymmetric quantization does not incur additional overhead during inferences.


\subsection{Previous Bit-Slice GEMMs}

\begin{figure}[!t]
    \centering
    \includegraphics{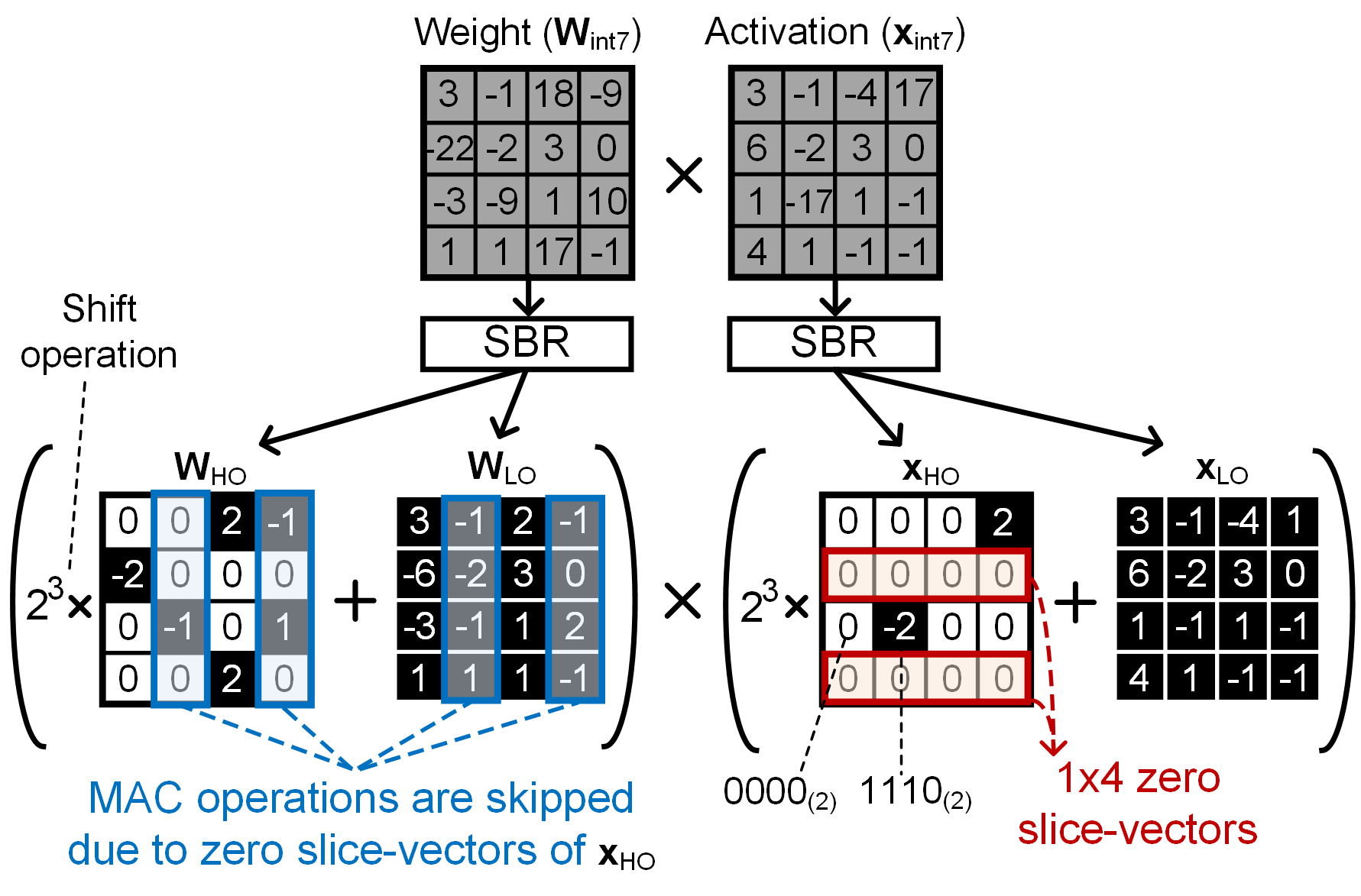}
    \vspace*{-5mm}
    \caption{An example of the bit-slice GEMM using 7-bit symmetric quantization and the SBR for both weight and activation\cite{im2024sibia}.}
    \vspace*{-2mm}
    \label{fig:sbr}
\end{figure}

To enhance hardware efficiency during dense model inferences, recent DNN accelerators \cite{han2023hnpu, shomron2020non, im2024sibia, im2024lutein} utilize low 
precision, and further mitigate the complexity of integer GEMMs by segmenting integers into bit-slices and leveraging sparsity at high-order (HO) slices.
For example, as illustrated in Fig. \ref{slicing}(a), the straightforward bit-slicing \cite{shomron2020non} typically divides an 8-bit integer into a 4-bit signed HO slice and a 4-bit unsigned low-order (LO) slice, skipping MAC operations for $0000_{(2)}$ HO slices.
However, this approach cannot skip $1111_{(2)}$ HO slices, which occur as frequently as $0000_{(2)}$ in symmetric quantization \cite{glorot2010understanding, he2015delving, lecun2002efficient, shridhar2019probact, neal2012bayesian}, limiting slice-level sparsity.

The signed bit-slice representation (SBR)\cite{im2024sibia}, as shown in Fig. \ref{slicing}(b), has been introduced to overcome this limitation. 
The SBR captures zero HO slices from both positive and negative near-zero values by dividing a $(3n+4)$-bit integer into one 4-bit signed HO slice and $n$ 3-bit unsigned LO slices, where $n$ is a positive integer.
Then, 3-bit unsigned LO slices are extended to 4-bit signed slices. 
For the case of $n=1$ depicted in Fig. \ref{slicing}(b), the 3-bit LO slice is first extended to a 4-bit signed slice by appending the sign bit of the HO slice, and then $\text{1111}_{(2)}$ HO slice is added by $\text{0001}_{(2)}$ to compensate for the attached sign bit, converting it to $\text{0000}_{(2)}$.
Given the enhanced slice-level sparsity, the bit-slice GEMM in \cite{im2024sibia} is designed to group $v$ slices into a $v$-length slice-vector and skip operations for vectors that consist entirely of zero slices.
More precisely, based on the SBR dividing a 7-bit integer value into two 4-bit slices \cite{im2024sibia}, the bit-slice GEMM is calculated as 
\begin{align}
    \textbf{W}_{\text{int7}}\textbf{x}_{\text{int7}} + \hat{\textbf{b}}_{\text{int32}} = (\textbf{W}_{\text{HO}} + \textbf{W}_{\text{LO}})(\textbf{x}_{\text{HO}} + \textbf{x}_{\text{LO}}) + \hat{\textbf{b}}_{\text{int32}}.
\end{align}
There are four bit-slice computations, i.e., $\textbf{W}_{\text{HO}}\textbf{x}_{\text{HO}}$, $\textbf{W}_{\text{LO}}\textbf{x}_{\text{HO}}$, $\textbf{W}_{\text{HO}}\textbf{x}_{\text{LO}}$, and $\textbf{W}_{\text{LO}}\textbf{x}_{\text{LO}}$, which are probably sparse GEMMs except for $\textbf{W}_{\text{LO}}\textbf{x}_{\text{LO}}$.
Note that the previous bit-slice GEMM only supports to skip the operations of zero-slice vectors related to either  $\textbf{W}_{\text{HO}}$ or $\textbf{x}_{\text{HO}}$.
Fig. \ref{fig:sbr} illustrates an example of the bit-slice GEMM using 7-bit symmetric quantization alongside SBR and 4-length slice-vectors, i.e., $v=4$.
The SBR and grouping slices into vectors \cite{im2024sibia} still exploit bit-slice sparsity and effectively reduce computational complexity.
However, since these methods only consider skipping operations of zero HO slices, it is impossible to exploit slice-level sparsity for asymmetric quantization, which generates few zero HO slices.    

\subsection{Motivation}

\begin{figure}
    \centering
    \vspace*{-3mm}
    \includegraphics{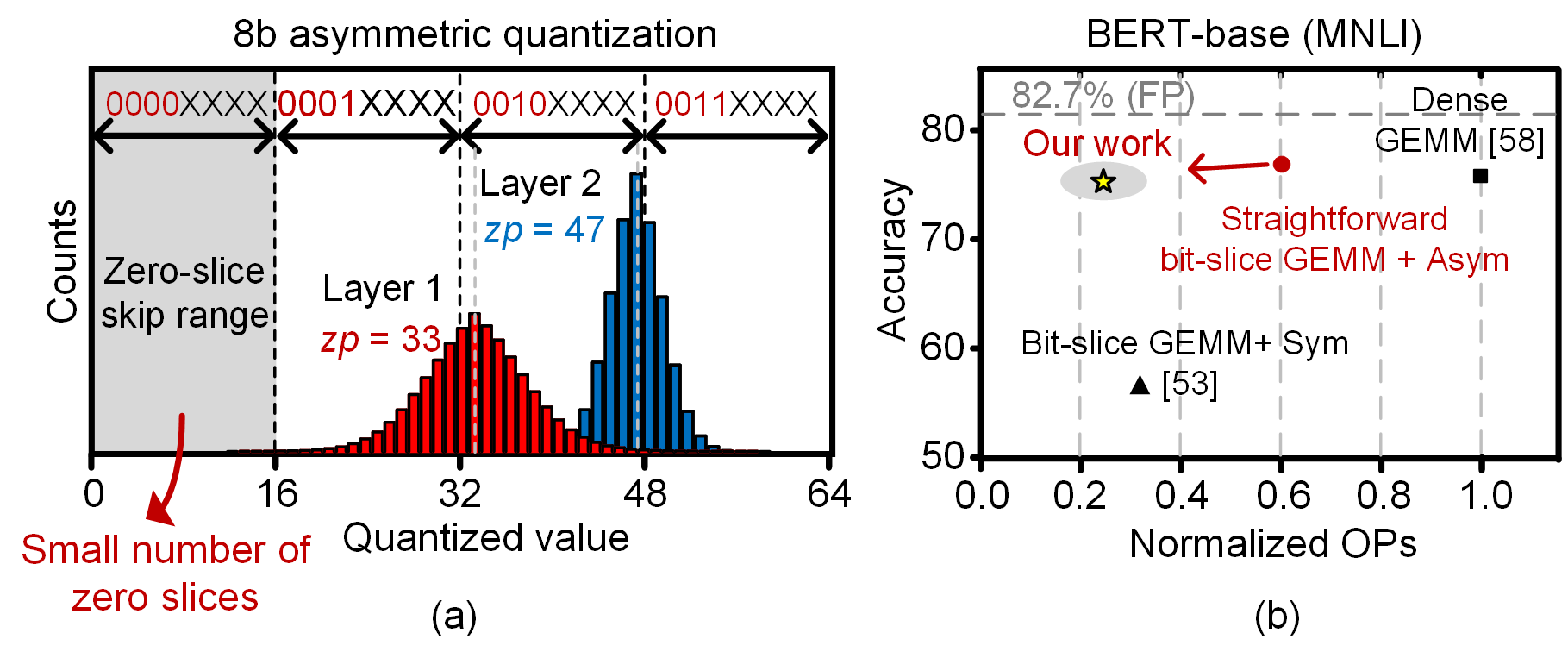}
    \vspace*{-5mm}
    \caption{(a) Distributions of asymmetrically quantized activations. (b) Accuracy comparison when using different GEMMs for BERT-base\cite{devlin2018bert} and GLUE dataset (MNLI) \cite{wang2018glue}. }
    \label{fig:asym_dist}
    \vspace*{-3mm}
\end{figure} 
Since most of the actual activations in DNN layers have an asymmetric distribution \cite{liu2023noisyquant, yuan2022ptq4vit}, symmetric quantization cannot fully utilize the quantization bit-width, potentially leading to accuracy loss.  
To address this challenge, recent algorithm-level works have embraced asymmetric quantization, specifically for activations \cite{lin2021fq, liu2023pd, lee2023flexround, cai2020zeroq, li2023repq, wei2022outlier, wei2023outlier, nagel2021white, liu2023qllm, shao2023omniquant}.
%
However, one drawback of asymmetric quantization is that it does not generate sufficient zero slices, which can be skipped in the existing bit-slice GEMMs \cite{im2024sibia, shomron2020non,han2023hnpu, im2024lutein}, as shown in Fig. \ref{fig:asym_dist}(a). 
To address this limitation, there is a clear need for a novel bit-slice GEMM approach, which skips frequent nonzero slices in asymmetrically-quantized activations, and optimization methods that increase the frequency of skippable nonzero slices.
In response to this need, this paper introduces a new bit-slice GEMM along with several optimization methods and its accelerator, \textit{Panacea}, to reduce memory accesses and the number of computations while maintaining high accuracy, as shown in Fig. \ref{fig:asym_dist}(b).

\section{Panacea Accelerator}
\subsection{Overview}

\begin{figure}[!t]
    \centering    
    \includegraphics{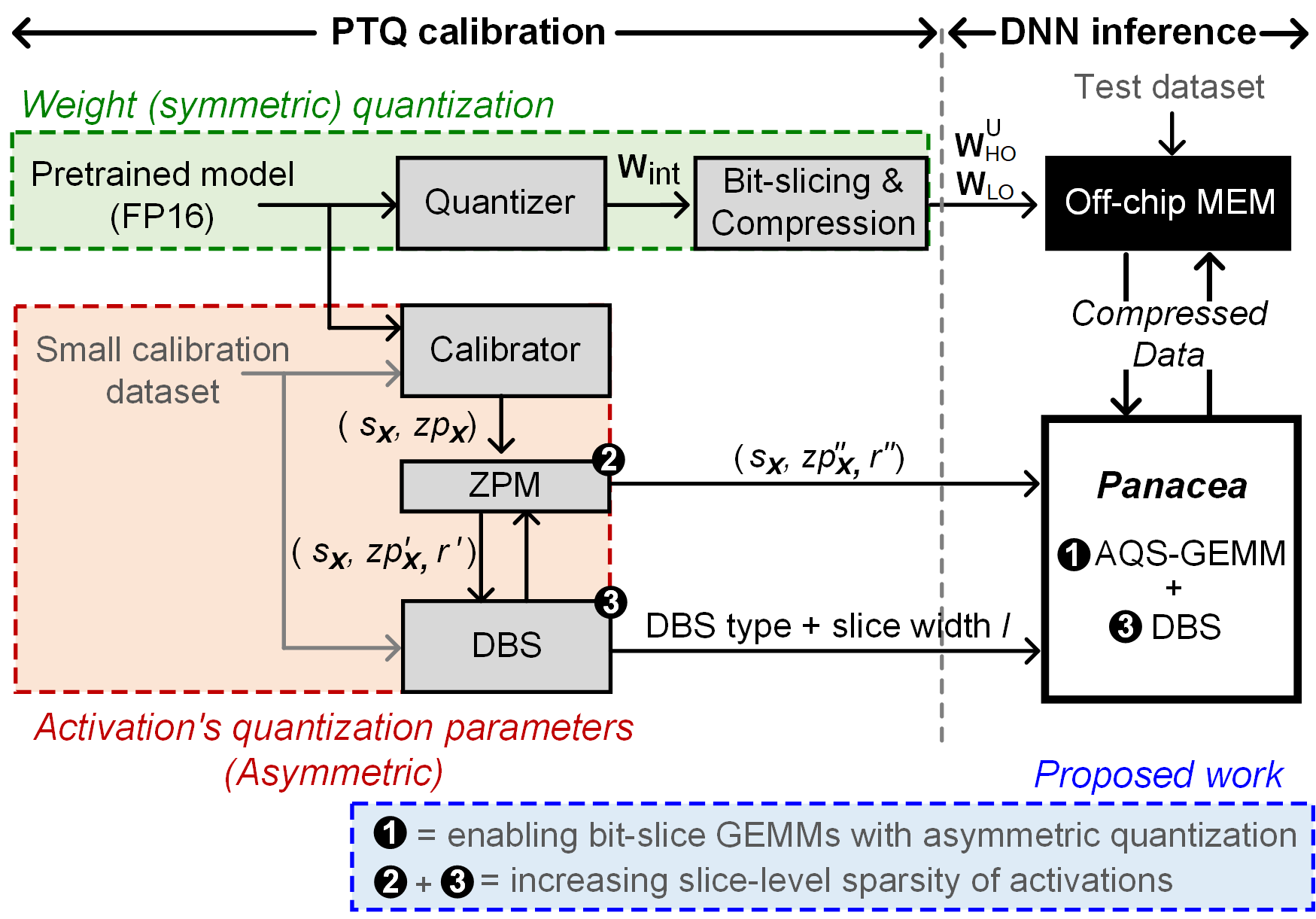}
    \caption{Overview of PTQ calibration and DNN inference, which are based on the proposed methods and \textit{Panacea}.}
    \vspace*{-3mm}
    \label{fig:overview}
\end{figure}

\textit{Panacea} is a novel DNN accelerator that supports both asymmetric quantization and sparse bit-slice computations to achieve high accuracy and energy efficiency.
Fig. \ref{fig:overview} shows an overview of the proposed work, mainly consisting of \circled{1} asymmetrically quantized bit-slice GEMM (AQS-GEMM) in \textit{Panacea}, which enables the bit-slice GEMM for asymmetric quantization during inference (Section III-B), and optimization methods: \circled{2} zero-point manipulation (ZPM) and \circled{3} distribution-based slicing (DBS), which increase slice-level sparsity of activations during calibration (Section III-C).
The PTQ calibration quantizes weights and extracts activation's quantization parameters, including $s_{\textbf{x}}$ and $zp_{\textbf{x}}$, as outlined in (1) and (2).
The parameters are adjusted through \circled{2} ZPM and \circled{3} DBS to enhance slice-level sparsity by considering the slice skip range of the AQS-GEMM.
During inference, \textit{Panacea} utilizes the compressed weights, symmetrically quantized during calibration, and compressed activations, asymmetrically quantized based on the parameter values obtained from calibration, for bit-slice GEMMs.
\circled{1} \textit{Panacea}'s AQS-GEMM core efficiently handles bit-slice computations by skipping not only zero slices in symmetric quantization, but also frequent nonzero slices in asymmetric quantization.
Note that accumulated GEMM results, i.e., activations, are re-quantized and compressed for the next layer.
In terms of hardware architecture, \textit{Panacea} separates dynamic and static workload operators to optimize sparse and dense computations, maximizing data reuse and minimizing EMA (Section III-D).


\begin{figure*}
    \centering
    \includegraphics{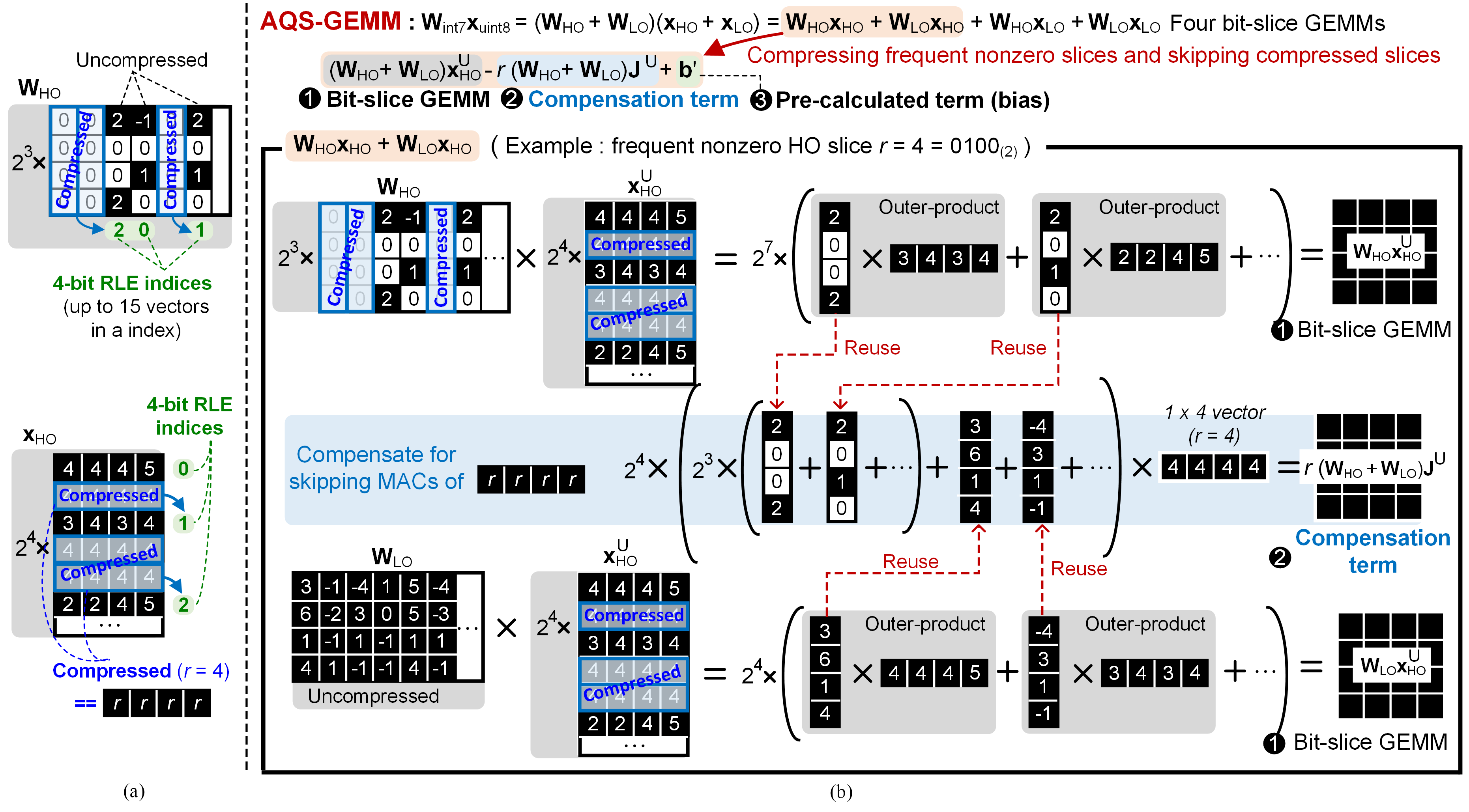}
    \caption{(a) Compressing vectors only comprising frequent slices in $\textbf{x}_{\text{HO}}$. (b) The concept of AQS-GEMM skipping MACs of compressed HO vectors.}
    \label{proposed_main}
\end{figure*}

\subsection{AQS-GEMM}
\textit{Panacea} implements symmetric quantization for weights and asymmetric quantization for activations. 
For weight quantization, it adopts SBR \cite{im2024sibia} segmenting ($3n+4$)-bit weight $\textbf{W}_{\text{int}}$ into ($n+1$) 4-bit sliced matrices, while for activation quantization, it utilizes the straightforward bit-slice representation method \cite{shomron2020non, sharma2018bit} segmenting ($4k+4$)-bit activation $\textbf{x}_{\text{uint}}$ into ($k+1$) 4-bit sliced matrices.
For a good understanding, this section explains the case of $n=1$ and $k=1$, i.e., $\textbf{W}_{\text{int7}}$ and $\textbf{x}_{\text{uint8}}$.
To compute $\textbf{W}_{\text{int7}}\textbf{x}_{\text{uint8}}$, \textit{Panacea} divides it into four bit-slice computations: $\textbf{W}_{\text{HO}}\textbf{x}_{\text{HO}}$,$\textbf{W}_{\text{LO}}\textbf{x}_{\text{HO}}$,$\textbf{W}_{\text{HO}}\textbf{x}_{\text{LO}}$, and $\textbf{W}_{\text{LO}}\textbf{x}_{\text{LO}}$.

As depicted in Fig. \ref{fig:asym_dist}(a), asymmetric quantization creates distributions that are not centered around zero, frequently observing specific nonzero HO slices near their zero point values ($zp$).
The floating $zp$ for different activations complicates the implementation of bit-slice GEMMs, as it changes a frequent nonzero HO slice observed in each layer.
To address these challenges and enhance computational efficiency, we propose the AQS-GEMM, which compresses out these frequent nonzero slices generated by asymmetric quantization, and skips their MAC operations.

\noindent\textbf{Compressing HO slices.} 
The AQS-GEMM operates under the assumption that HO slices are sparse and LO slices are dense.
As illustrated in Fig. \ref{proposed_main}(a), the AQS-GEMM applies grouping and compression to HO slices.
First, it groups $\textbf{W}_{\text{HO}}$ into 4$\times$1 slice vectors and $\textbf{x}_{\text{HO}}$ into 1$\times$4 slice vectors.
Then, the run-length encoding (RLE)\cite{im2024sibia} is applied to the slice vectors.
The RLE index bit-width denotes how many successive vectors can be compressed at a time. 
In the case of 4-bit RLE indices, we can compress up to 15 successive 1$\times$4 vectors into an index.
More specifically, in the case of symmetrically-quantized $\textbf{W}_{\text{HO}}$, a slice-vector is compressed if all slices within the vector have zero values.
For the asymmetrically-quantized $\textbf{x}_{\text{HO}}$, a slice-vector is compressed if all slices in the vector are $r$-valued slices, where $r$ is an HO slice of 8-bit zero point, i.e., $r = zp_{\text{HO}}$.
Note that many quantized values are observed around $zp$ due to the asymmetric quantization we applied. This results in $r$-values becoming predominant and others being sparse within the HO slices. 
With this approach, we exploit high HO vector sparsity in both weights and activations across numerous DNN layers, reducing EMA and the number of accesses from SRAM to the $\textit{Panacea}$'s processing engine.
Our evaluation with DNN benchmarks, specifically for DeiT-base\cite{touvron2021training} on the ImageNet-1k dataset \cite{deng2009imagenet} and GPT-2 \cite{radford2019language} on the WikiText-2\cite{wang2018glue}, demonstrates that compressing HO slices reduces EMA by 60.5\% and 46.8\%, respectively, compared to the baseline \cite{im2024sibia} using uncompressed data format from DRAM to the processing core.
Furthermore, the AQS-GEMM's compression reduces SRAM access by 29.2\% for DeiT-base and 27.4\% for GPT-2, compared to \cite{im2024sibia}.
These reductions improve energy efficiency and mitigate the memory bandwidth demands of large-scale DNNs.
In a later section, we further enhance the effect of compression by increasing slice-level sparsity followed by high vector-level sparsity.

\noindent\textbf{Skipping compressed HO slice vectors.}
As discussed above, the $\textbf{W}_{\text{HO}}$'s slice-vectors that only have zero slices, and the $\textbf{x}_{\text{HO}}$'s slice-vectors that only have $r$-value are compressed out.
This compression saves computation by skipping operations involving these compressed vectors, and it focuses computation only on the uncompressed vectors to optimize hardware efficiency.
The proposed AQS-GEMM executes four bit-slice GEMMs, newly presenting a method skipping the compressed slices in the asymmetrically-quantized $\textbf{x}_{\text{HO}}$ as shown in Fig. \ref{proposed_main}(b).
In contrast to directly skipping zero-valued slices in the previous bit-slice GEMM \cite{im2024sibia, shomron2020non, han2023hnpu, im2024lutein}, $r$-valued slices in $\textbf{x}_{\text{HO}}$ cannot be explicitly skipped to compute exact results. 
To skip the compressed $r$-valued slices and get exact GEMM results simultaneously, the proposed method reformulates $(\textbf{W}_{\text{HO}} + \textbf{W}_{\text{LO}})\textbf{x}_{\text{HO}}$ as follows.
\begin{align}
    & (\textbf{W}_{\text{HO}} + \textbf{W}_{\text{LO}})\textbf{x}_{\text{HO}} \notag \\
    & = (\textbf{W}_{\text{HO}} + \textbf{W}_{\text{LO}})(\textbf{x}^{\text{U}}_{\text{HO}} + \textbf{x}^{\text{C}}_{\text{HO}}) \notag \\
    & = (\textbf{W}_{\text{HO}} + \textbf{W}_{\text{LO}})(\textbf{x}^{\text{U}}_{\text{HO}} + r \times \textbf{J}^{\text{C}}) \\
    & = (\textbf{W}_{\text{HO}} + \textbf{W}_{\text{LO}})(\textbf{x}^{\text{U}}_{\text{HO}} + r \times (\textbf{1}^{K \times N}-\textbf{J}^{\text{U}})) \notag \\
    & = (\textbf{W}_{\text{HO}} + \textbf{W}_{\text{LO}})\textbf{x}^{\text{U}}_{\text{HO}} - r(\textbf{W}_{\text{HO}} + \textbf{W}_{\text{LO}})\textbf{J}^{\text{U}} + \textbf{b}',
\end{align}
where $\text{C}$ indicates the set of compressed slices, $\text{U}$ indicates the set of uncompressed slices, $\textbf{x}^{\text{C}}_{\text{HO}}$ and $\textbf{x}^{\text{U}}_{\text{HO}}$ are subset matrices of $\textbf{x}_{\text{HO}}$ only including compressed and uncompressed slices, respectively, and $\textbf{J}^{\text{C}}_{\text{HO}}$ and $\textbf{J}^{\text{U}}_{\text{HO}}$ are matrices with 1 at indices of compressed and uncompressed slices, respectively.
Note that $\textbf{x}_{\text{HO}} = \textbf{x}^{\text{U}}_{\text{HO}} + \textbf{x}^{\text{C}}_{\text{HO}}$, $\textbf{x}^{\text{C}}_{\text{HO}} = r \times \textbf{J}^{\text{C}}$ and $\textbf{J}^{\text{C}} =  \textbf{1}^{K\times N} - \textbf{J}^{\text{U}}$.

In (5), the case of symmetric quantization ($r = 0$)  \cite{im2024sibia, shomron2020non, han2023hnpu, im2024lutein} becomes equivalent to $(\textbf{W}_{\text{HO}} + \textbf{W}_{\text{LO}})\textbf{x}^{\text{U}}_{\text{HO}}$, which is only associated with the activation's uncompressed slices. 
However, in the case of asymmetric quantization $(r \neq 0)$, a compensation term $(\textbf{W}_{\text{HO}} + \textbf{W}_{\text{LO}})(r \times \textbf{J}^{\text{C}})$ should be added to compute the exact GEMM compared to the symmetric quantization.
Unfortunately, the naive computation of compensation term requires additional memory accesses to load the weight slices associated with $\textbf{J}^{\text{C}}$, which are originally skipped due to compressing $\textbf{x}_{\text{HO}}$ slices.
The loaded weight slices for the compensation term cannot be reused, and the overhead is further exacerbated due to the increased number of compressed vectors in the case of high slice sparsity.
To minimize the overhead, we derive (6) from (5), calculating the compensation term with weight slices only associated with the uncompressed slices $(\textbf{x}^{\text{U}}_{\text{HO}})$. 
This enables the reuse of weight slices loaded for the bit-slice computations $(\textbf{W}_{\text{HO}} + \textbf{W}_{\text{LO}})\textbf{x}^{\text{U}}_{\text{HO}}$, as shown in Fig. \ref{proposed_main}(b), accumulating the loaded weight slices and then performing an outer product with the 1$\times$4 vector containing only $r$.
Meanwhile, $\textbf{b}'$ is pre-computed offline by $(\textbf{W}_{\text{HO}} + \textbf{W}_{\text{LO}})(r \times \textbf{1}^{K \times N})$, and it is added to the layer's bias in advance.
As a result, with negligible overhead, the AQS-GEMM efficiently computes the bit-slice GEMM and the compensation term for exact results.
Note that after decoding RLE indices to recover the original vector indices, as shown in Fig. \ref{proposed_main}(b), the bit-slice GEMM of AQS-GEMM is executed by outer products with pairs of uncompressed 4$\times$1 weight and 1$\times$4 activation slice-vectors that have matching indices, each of which produces 4$\times$4 partial-sums.

\begin{table}[]
\caption{Hardware workloads in bit-slice GEMM accelerators}
\vspace*{-1.5mm}
\resizebox{\columnwidth}{!}{
\renewcommand{\arraystretch}{1.3}
\begin{tabular}{c|c|c|cc}
\hline 
Accel. & \textit{Sibia}\cite{im2024sibia} & \multicolumn{3}{c }{\textit{Panacea} (AQS-GEMM core)} \\
\hline 
Core's & \multirow{2}{*}{Bit-slice GEMMs} &  {Bit-slice GEMMs} & \multicolumn{2}{c }{Compensation} \\  \cline{4-5}
       comput.&  &  {w/o compensation} & \multicolumn{1}{c|}{ In (5) }  & In (6)  \\ \hline
Mul.  & $32K(2-\text{max}(\rho_{\text{x}}, \rho_{\text{w}}))$ &  {$16K(2-\rho_{\text{x}})(2-\rho_{\text{w}})$}  & \multicolumn{2}{c }{16}   \\ \hline
Add.    & $32K(2-\text{max}(\rho_{\text{x}}, \rho_{\text{w}}))$ &   $16K(2-\rho_{\text{x}})(2-\rho_{\text{w}})$  & \multicolumn{1}{c|}{$8K\rho_{\text{x}}$}  & $8K(1-\rho_{\text{x}})$    \\ \hline
EMA  & $ 14K  $  & $ 4K(4-\rho_{\text{w}}-\rho_{\text{x}})  $  & \multicolumn{1}{c|}{ $8K\rho_{\text{x}}$ }      & 0   \\ \hline
\end{tabular}
}
\vspace*{-3.5mm}
\label{table1}
\end{table}

\noindent
\textbf{Formalization of hardware workloads.} 
Table \ref{table1} shows the relation between sparsity and hardware workloads, such as 4b$\times$4b multiplications, 8b additions, and 4b EMA, in two bit-slice accelerators with sufficient memory space assumed.
The workloads are formulized by the example using two bit-slices for both weights $(\mathbf{W}_{\text{int7}} \in \mathbb{Z}^{4 \times K})$ and activations $(\mathbf{x}_{\text{int7}} \in \mathbb{Z}^{K \times 4}$ for \textit{Sibia} and $\mathbf{x}_{\text{uint8}} \in \mathbb{N}^{K \times 4}$ for \textit{Panacea}$)$.
While \textit{Sibia}\cite{im2024sibia} only supports one of the HO vector-level sparsities of weights and activations ($\rho_\text{w}, \rho_\text{x}\leq 1$), \textit{Panacea} accomodates both sparsities.
Note that in \textit{Sibia}, which uses symmetrically quantized activations, $\rho_\text{x}$ represents the ratio of all-zero slice vectors, whereas in \textit{Panacea}, it represents a broader range that includes $r$-valued slice vectors.
Although \textit{Panacea} requires an additional compensation term for exact computations unlike \textit{Sibia}\cite{im2024sibia}, the compensation term incurs only 16 extra multiplications and low-cost additions that result in a negligible hardware overhead.
In terms of EMA for bit-slice GEMMs, \textit{Panacea} loads only the uncompressed parts, leading to lower EMA compared to \textit{Sibia} \cite{im2024sibia}, which loads dense matrices regardless of compression. 
The naive $8K\rho_\text{x}$ EMA overhead of the required compensation term in (5) can be eliminated by the transition to (6), which reuses the weight slices corresponding to $\textbf{x}_\text{HO}^{\text{U}}$ already loaded for bit-slice GEMMs.
Thus, the AQS-GEMM core in \textit{Panacea} fully leverages the sparsities for exact computations, optimizing both computational and memory efficiency.


\noindent
\textbf{Scalability for lower bit-precisions.}
The AQS-GEMM is scalable to support any bit-width format following the $(3n+4)$-bit weight and $(4k+4)$-bit activation format, thereby supporting lower-bit quantization, such as 4-bit weights when $n=0$.
It is also possible to extend the idea of AQS-GEMM to either lower-bit slices.
Although recent quantization works, such as OPTQ\cite{frantar2022optq} and AWQ \cite{lin2024awq}, have recently studied 4-bit or lower bit-precisions, they work only for weights, using 16-bit FP numbers for activations and assuming 16-bit FP multipliers. 
Thus, when considering low-bit quantization for both weights and activations, the 8-bit precision is still an attractive option to achieve high accuracy and hardware efficiency simultaneously.

\subsection{Enhancing Slice-Level Sparsity}
Since the AQS-GEMM achieves better efficiency thanks to the increased number of compressed vectors, increasing slice-level sparsity for both weights and activations is important.
Note that improving the slice-level sparsity generally results in increased vector-level sparsity.
However, activation's asymmetric quantization produces varying distributions, leading to low slice sparsity.
To address this problem, we propose two algorithm-hardware co-optimization methods: zero-point manipulation (ZPM) and distribution-based slicing (DBS) within the PTQ calibration depicted in Fig. \ref{fig:overview}.


\begin{figure}[!t]
    \centering
    \includegraphics{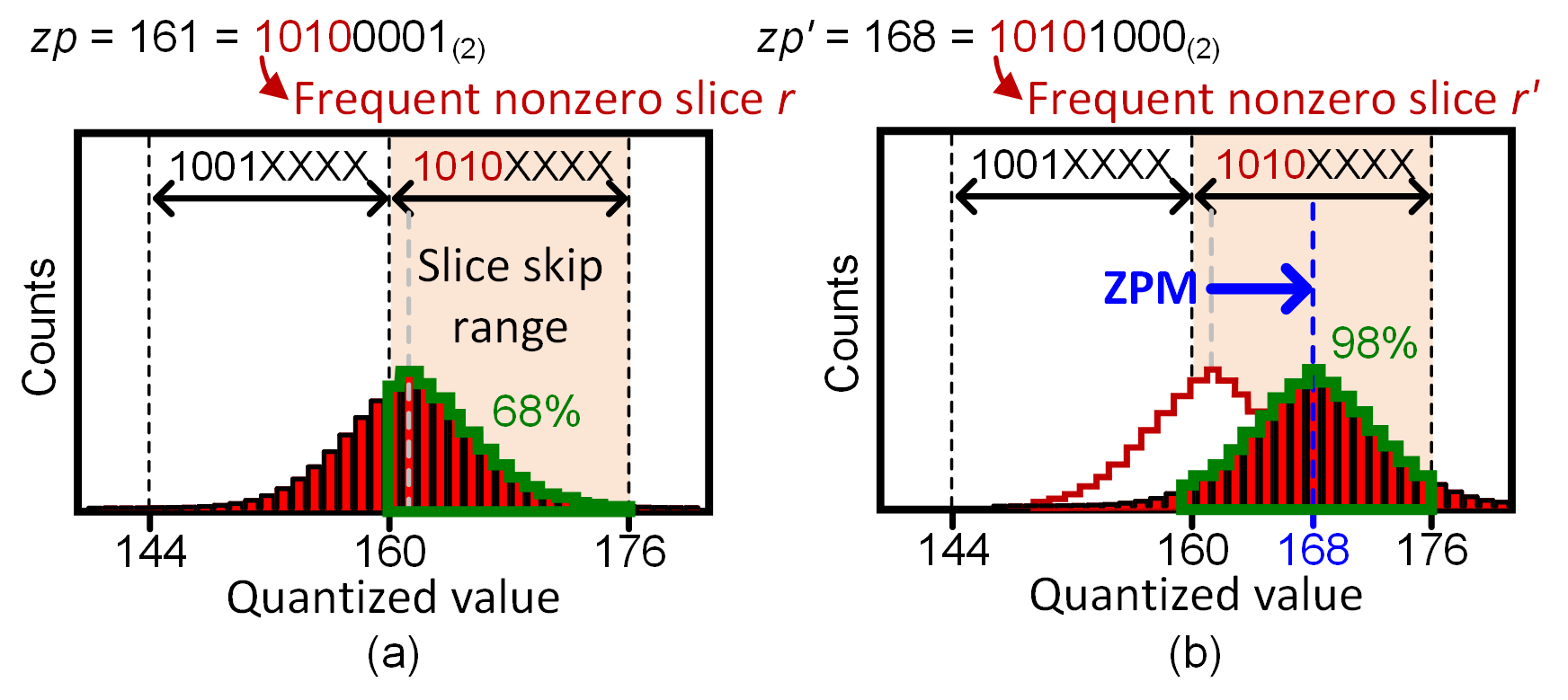}
    \vspace*{-3mm}
    \caption{Distributions of an asymmetrically quantized activation (a) without the ZPM, and (b) with the ZPM.}
    \label{fig:zpm}
    \vspace*{-3mm}
\end{figure}

\noindent\textbf{Sparsity-aware zero-point manipulation.}
Fig. \ref{fig:zpm}(a) shows an example of asymmetric quantization leading to low sparsity at HO slices of an input activation.
The example activation distribution of an FC layer in OPT-2.7B \cite{zhang2022opt} is centered around the zero point $zp=161$. 
When applying the compression of HO slices, the frequent HO slice becomes $r = 1010_{(2)}$, and only about 68\% of total values are present in the skip range observing $1010_{(2)}$ HO slices.
The actual compression ratio of $\textbf{x}_{\text{HO}}$ slices would be significantly lower than 68\% due to grouping slices into 1$\times$4 vectors before compression, leading to low efficiency in the AQS-GEMM. 
To address this problem, the ZPM increases the number of compressible HO slices by adjusting the zero point during the PTQ's calibration as
\begin{align}
    zp' = 
    \begin{cases}
        2^l\left\lfloor{zp / 2^l} \right\rfloor + 2^{l-1}, & \mbox{if } zp>0 \\
        0, & \mbox{otherwise },
    \end{cases}
\end{align}
where $l$ indicates the bit-width of LO slice, and $zp'$ indicates a manipulated zero-point.
After adjusting the zero point, the frequent HO slice is also changed by $r' =(zp'-2^{l-1})_{\text{HO}}$.
Note that $\textit{Panacea}$ uses $l=4$ to divide an 8-bit value into two 4-bit slices.
Fig. \ref{fig:zpm}(b) shows the shifted distribution by the ZPM. 
The center of distribution $(zp')$ matches with the center of skip range, thereby putting more values in the skip range.
The slice-level sparsity increases from 68\% to 98\%.
The detailed slice-sparsity analysis on actual DNN benchmarks will be shown in Section IV.
By increasing slice sparsity with the ZPM, the AQS-GEMM further reduces the number of operations by 33\% for the FC layer of OPT-2.7B \cite{zhang2022opt}, compared to the AQS-GEMM without the ZPM.
Note that we observed the slight distribution shift of the ZPM does not cause a considerable change in accuracy.

\noindent\textbf{Distribution-based slicing.}
Although the ZPM successfully increases sparsity, the distribution of quantized values in some layers spreads over a wide range, resulting in low slice sparsity since fewer quantized values fall within the skip range.
\begin{figure}[t]
    \centering
    \includegraphics{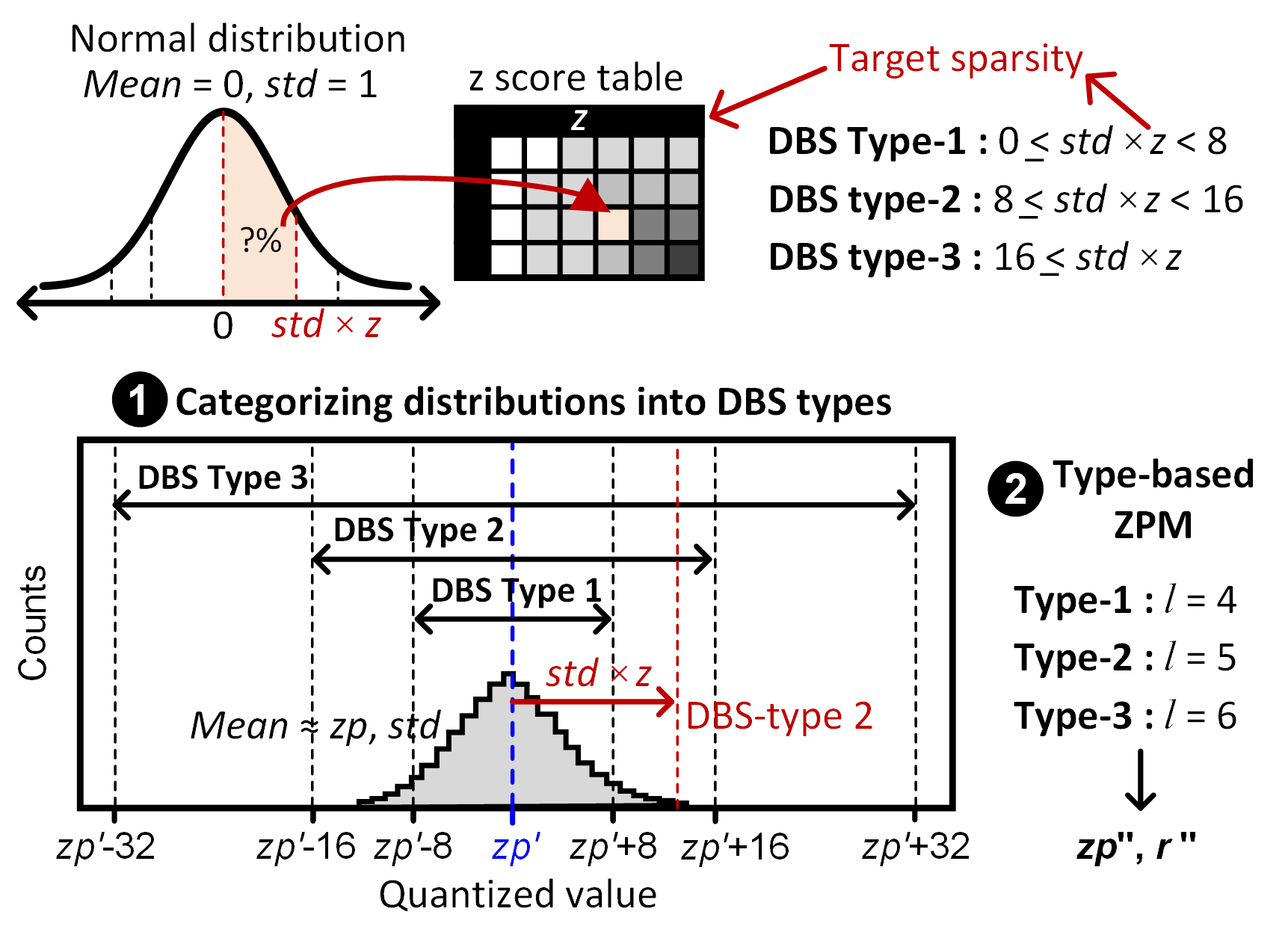}
    \caption{Categorizing distributions of quantized activations into three types based on their standard deviations, $z$ score table, and type-based ZPM.}
    \label{fig:DBS_type}
    \vspace*{-2.5mm}
\end{figure}
To enhance the slice sparsity of wide distributions, we newly propose DBS, which allows different slice's bit-widths depending on each activation distribution, enabling dynamic slice-skip ranges during inferences.
During the calibration shown in Fig. \ref{fig:overview}, the DBS quantitatively analyzes the activation distribution for each layer and categorizes it into three types to determine the slice's bit-width.
More precisely, the distribution monitoring step records histograms for quantized activations and then calculates their standard deviations ($std$).
Then, as depicted in Fig. \ref{fig:DBS_type}, \circled{1} Based on the $std$ and $z$-score table, which contains the distribution's area from the mean up to a given $std \times z$, we categorize distributions into three types by comparing $std \times z$ to three distinct ranges.
The DBS type-1 means the slice sparsity is originally high, and type-2 or 3 means the observed sparsity is lower than our target sparsity.
To increase the slice sparsity for type-2 and type-3, the DBS allocates fewer bits to the HO slice and more bits to the LO slice ($l>4$) than the basic scheme ($l=4$).
More precisely, we allocate $l=4$, $5$, and $6$ bits to the LO slice for type-1, type-2, and type-3, respectively, while we allocate $(8-l)$-bits to the HO slice for each type.
This dynamic bit slicing will expand their slice-skip ranges to achieve our target sparsity at inference.
\circled{2} Note that our calibration step employs a type-based ZPM by computing $zp''$ and $r''$ based on the modified slice's bit-width $l$ and (7) to maximize the slice sparsity.

\begin{figure}[t]
    \centering
    \includegraphics{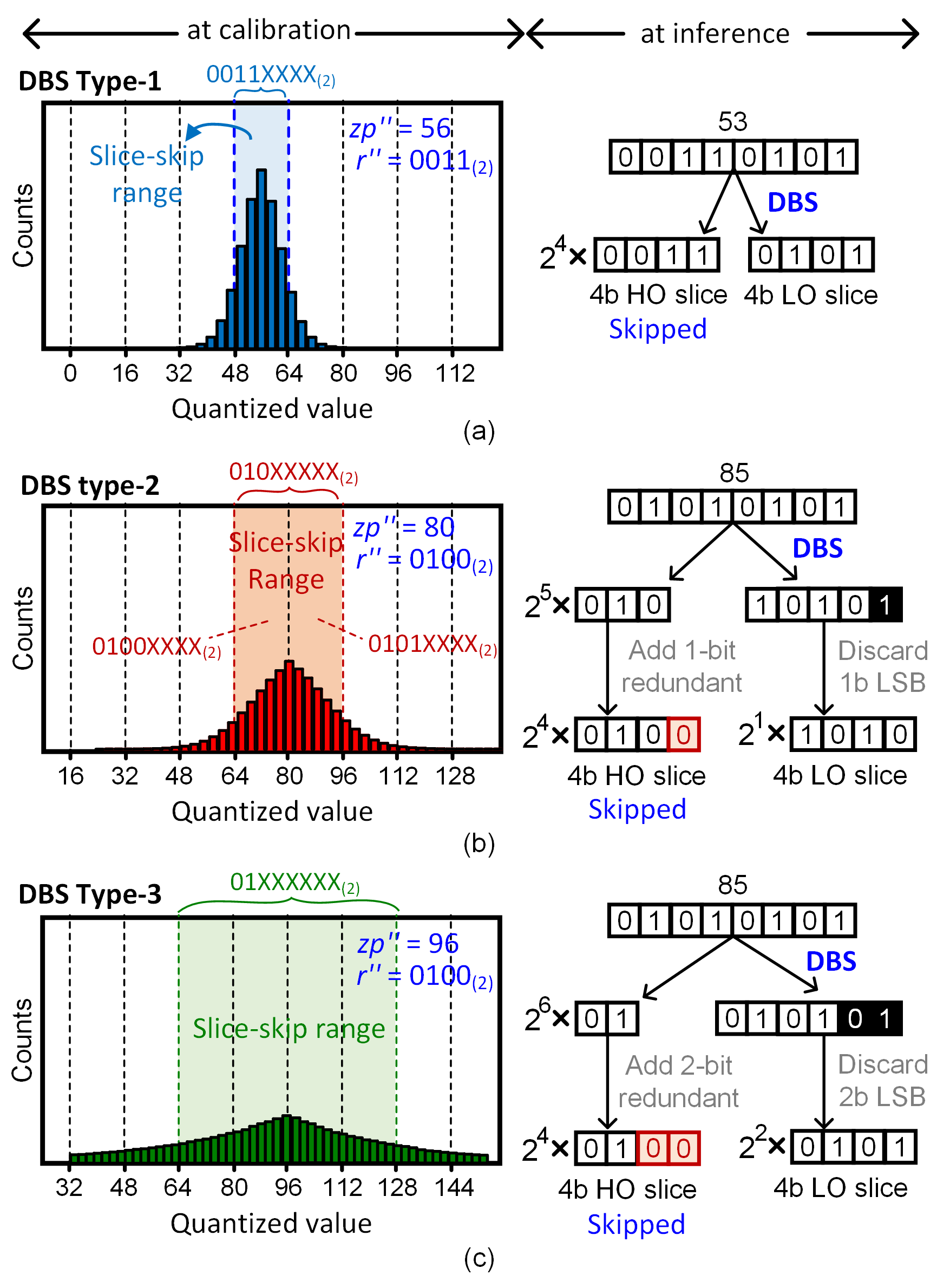}
    \vspace*{-3mm}
    \caption{Dynamically applying bit-slicing rules to different types during the inference phase: (a) DBS type-1, (b) DBS type-2, and (c) DBS type-3.}
    \label{fig:DBS_slice}
    \vspace*{-3mm}
\end{figure}

\begin{figure*}[t]
    \centering
    \includegraphics{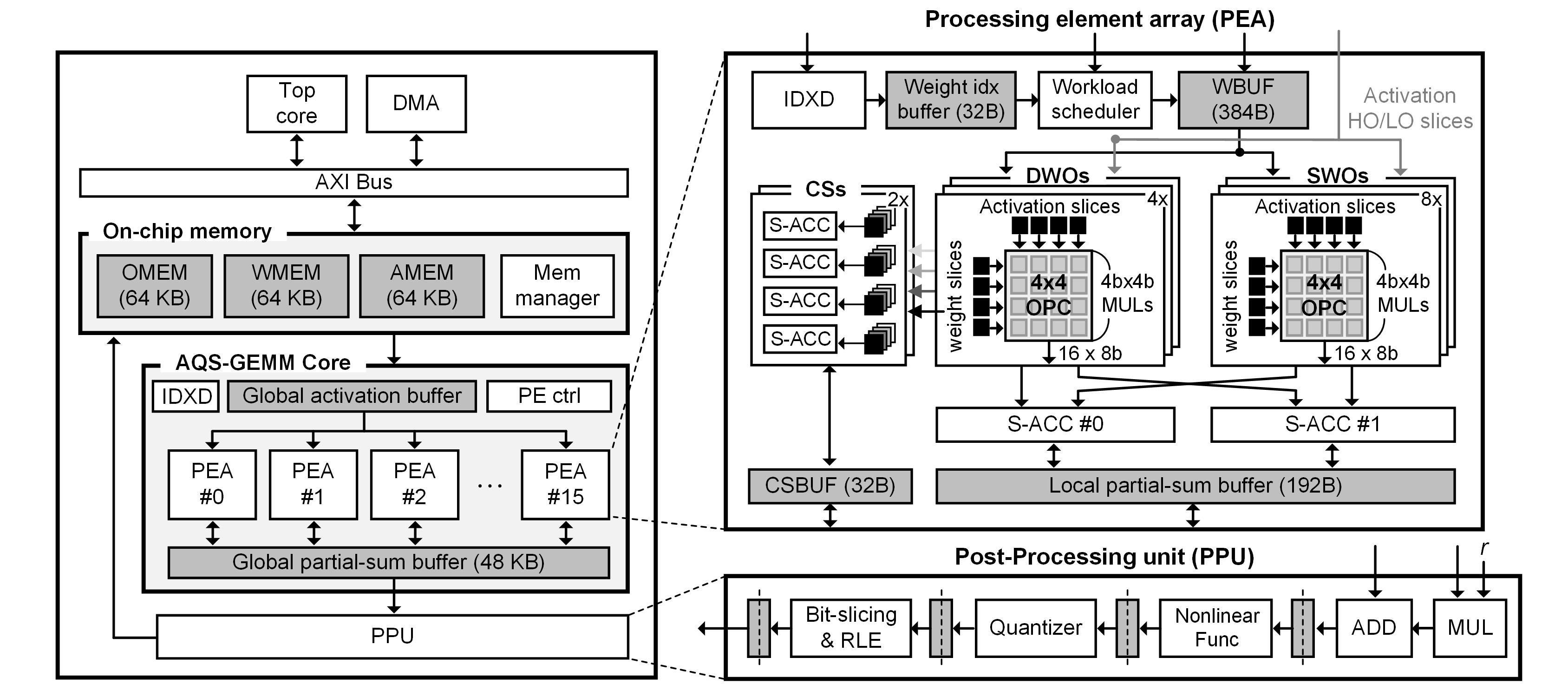}
    \caption{Overall architecture of \textit{Panacea} incorporating the AQS-GEMM core with 16 PEAs, each of which consists of 4 DWOs and 8 SWOs.}
    \label{fig:overall_architecture}
    \vspace*{-2mm}
\end{figure*}

\begin{figure}[t]
    \centering
    \vspace*{-2mm}
    \includegraphics{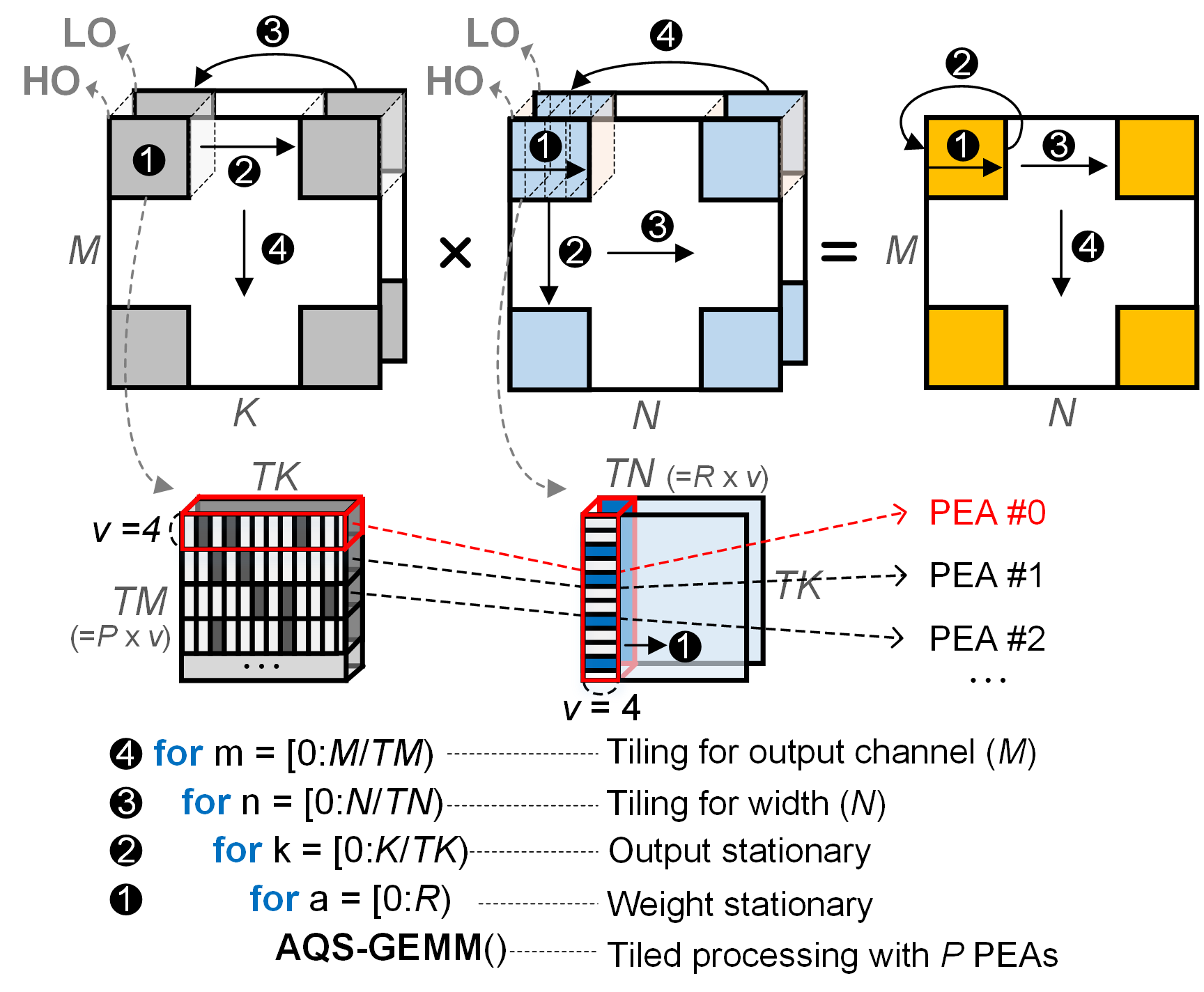}
    \vspace*{-2mm}
    \caption{\textit{Panacea}'s output stationary dataflow computing the tiled AQS-GEMM to maximize data reuse.}
    \vspace*{-4mm}
    \label{fig:dataflow}
\end{figure}

\begin{figure*}[t]
    \centering
    \includegraphics{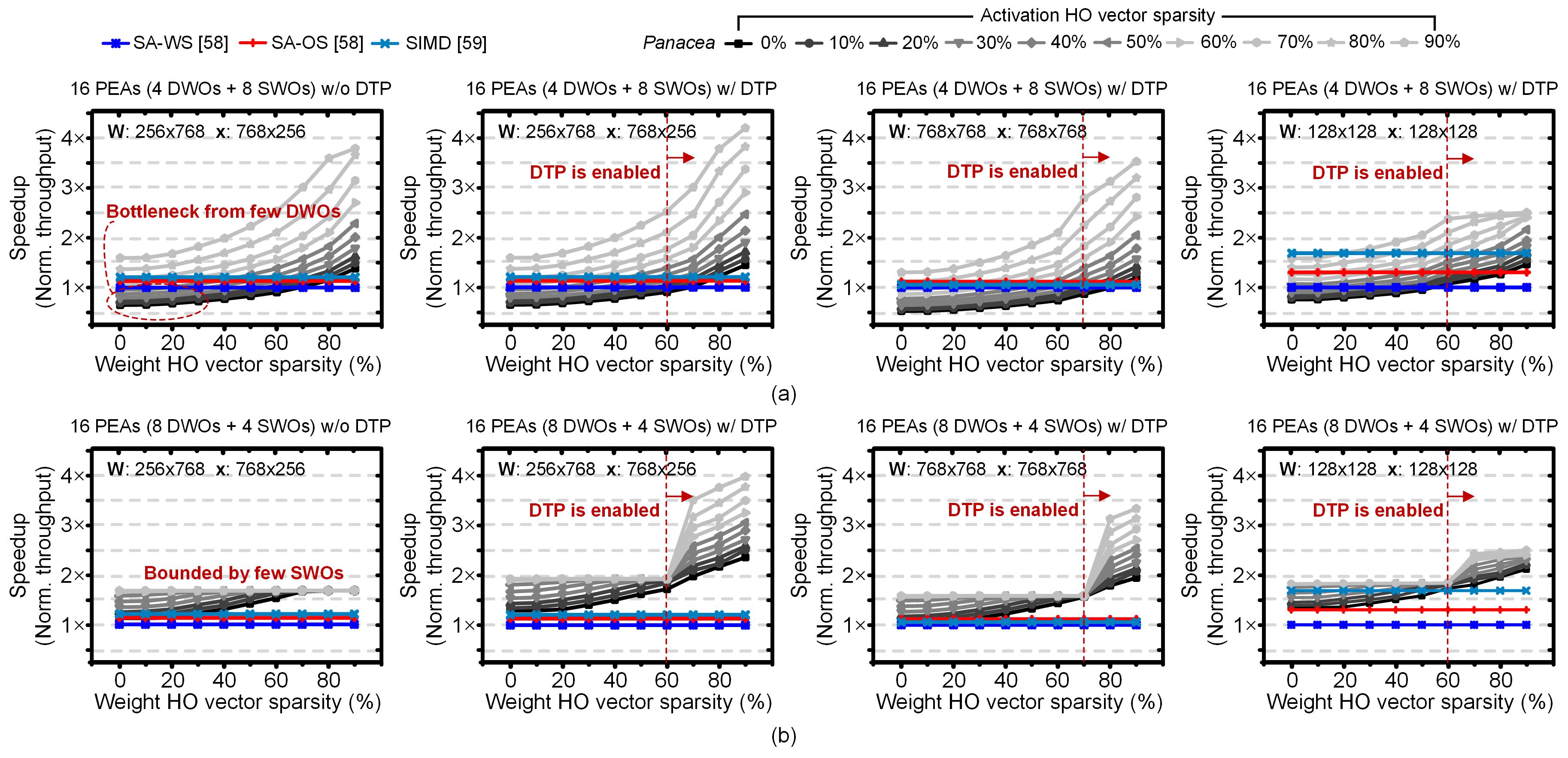}
    \vspace*{-5mm}
    \caption{Throughput evaluation of $\textit{Panacea}$ when using different design options, changing the HO slice sparsities of weight and activation, and changing the size of weight and activation; (a) 4 DWOs and 8 SWOs per PEA, and (b) 8 DWOs and 4 SWOs per PEA.}
    \vspace*{-2mm}
    \label{fig:design_option_eval}
\end{figure*}

In the inference step, given the identified DBS types and $l$, the DBS applies different slicing rules to compress and skip HO slices, as depicted in Fig. \ref{fig:DBS_slice}.
For example, as shown in Fig. \ref{fig:DBS_slice}(b), type-2 divides the 8-bit integer $\text{01010101}_{(2)}$ into $\text{010}_{(2)}$ HO slice and $\text{10101}_{(2)}$ LO slice, thereby resulting in the effect of expanding the skip range by two times, thanks to observing more values in the range for $\text{010}_{(2)}$ HO slices compared to the range for $\text{0101}_{(2)}$.
However, dynamically changing the slice bit width introduces significant hardware overhead.
To prevent the issue, as depicted in Fig. \ref{fig:DBS_slice}, DBS maintains the 4-bit width of slices by filling short HO slices with redundant zero bits and discarding LSBs in long LO slices.
This increases the average slice sparsity by 20\% (more than 50\% for some layers) while causing an acceptable accuracy loss of 0.6\%p on DeiT-base \cite{touvron2021training}.
In terms of hardware, the DBS is simply implemented by properly shifting the outputs of AQS-GEMM, resulting in a small overhead.
Consequently, the proposed DBS enables high slice sparsity in both narrow and wide distributions of activations, reducing EMA and maximizing the efficiency of AQS-GEMM in \textit{Panacea}.
The detailed evaluation results will be provided in Section IV.

\subsection{Hardware Design}
\noindent\textbf{Overall architecture.}
Fig. \ref{fig:overall_architecture} shows the overall architecture of $\textit{Panacea}$, which consists of a top controller, on-chip memory, an AQS-GEMM core, and a post-processing unit (PPU). 
The on-chip memory is divided into weight memory (WMEM), activation memory (AMEM), output memory (OMEM), and a memory manager. 
The AQS-GEMM core, with a global activation buffer and 16 processing element arrays (PEAs), efficiently computes the AQS-GEMM with weights and activations from the on-chip memory.
Each PEA has an index decoder (IDXD), a weight index buffer, a workload scheduler, a weight buffer (WBUF), dynamic workload operators (DWOs), static workload operators (SWOs), two compensators (CSs), a compensation buffer (CSBUF), a local partial-sum buffer and two shift-and-accumulator units (S-ACCs).
With RLE indices, the IDXD recovers the indices of uncompressed vectors and stores them in the weight index buffer.
By matching the decoded weight indices with indices of activation vectors received from the global buffer, the workload scheduler allocates outer products for uncompressed slices into DWOs and SWOs.
Each of DWO and SWO is comprised of an outer product calculator (OPC), which consists of 16 4b$\times$4b sign-unsigned multipliers to compute an outer product with a $4\times1$ weight slice-vector and a $1\times4$ activation slice-vector.
 
DWOs dynamically process $\textbf{W}_{\text{HO}}\textbf{x}_{\text{HO}}$, $\textbf{W}_{\text{LO}}\textbf{x}_{\text{HO}}$, and $\textbf{W}_{\text{HO}}\textbf{x}_{\text{LO}}$, while SWOs only handle the workload of dense operations, i.e., $\textbf{W}_{\text{LO}}\textbf{x}_{\text{LO}}$.
Based on the current layer's DBS type, each S-ACCs properly shifts partial sums calculated by DWOs and SWOs, and accumulates them.
Each CS consists of four small S-ACCs, which compute the compensation term of (6) by reusing the loaded weight slices.
The PPU adds the bit-slice computation outputs and CSs outputs for exact GEMM calculations.
After that, the PPU performs a non-linear function with the piecewise linear approximation, quantization, bit-slicing, HO slice compression, and RLE.
The outputs of PPU, i.e., uncompressed slices, are stored in OMEM.

\noindent\textbf{Dataflow.}
Fig.12 shows the dataflow of \textit{Panacea} and its loop representation, illustrating how it manages data movement to enhance efficiency.
$\textit{Panacea}$ supports an output stationary, eliminating the need for data movement of partial sums to external memory. 
This dataflow significantly reduces the latency and energy consumption of accessing external memory.
To maximize the reuse of weight, $\textit{Panacea}$ stores slices included in the $TM \times K$ weight tile into WMEM at once, if possible.
$\textit{Panacea}$ computes the tiled AQS-GEMM, as depicted in Fig. \ref{fig:dataflow}.
When the AQS-GEMM core processes a $TM \times TK$ weight tile along with a $TK \times TN$ activation tile, it assigns a $v \times TK$ weight sub-tile to each PEA, where each sub-tile includes both HO and LO slices, $P$ indicates the number of PEAs, $v$ indicates the length of slice vector, and $TM = Pv$.
This allocation ensures that the workload is evenly distributed among the PEAs for efficient parallel processing.
Furthermore, all PEAs share the $TK \times TN$ activation tile stored in the global activation buffer of the AQS-GEMM core, minimizing the repeated data load and removing the need for buffers storing the same activation tile.
Each PEA reuses the $v \times TK$ weight sub-tile by $R$ times, computing the tiled AQS-GEMM with HO and LO slices included in the $v \times TK$ weight sub-tile and $TK \times v$ activation sub-tile.
The tiled AQS-GEMM successfully maximizes the data reuse of sub-tiles by allocating both HO sub-tiles $(\textbf{W}_{\text{HO}}$ and $\textbf{x}_{\text{HO}})$ and LO sub-tiles $(\textbf{W}_{\text{LO}}$ and $\textbf{x}_{\text{LO}})$ to a PEA and computing $\textbf{W}_{\text{HO}}\textbf{x}_{\text{HO}}$,$\textbf{W}_{\text{LO}}\textbf{x}_{\text{HO}}$,$\textbf{W}_{\text{HO}}\textbf{x}_{\text{LO}}$, and $\textbf{W}_{\text{LO}}\textbf{x}_{\text{LO}}$ in the PEA.
This paper uses $v=4$, $P=16$, $TM=64$, $TK=32$, $TN=64$, and $R=16$.

\noindent\textbf{Double tile processing.}
At a high HO slice sparsity, $\textit{Panacea}$ cannot fully utilize the space of on-chip memory and experiences low utilization of DWOs.
To address the issue, we propose a double tile processing (DTP) that allows two different weight sub-tiles to be allocated into a PEA.
The DTP is enabled if WMEM can store slices included in the $2TM \times K$ weight tile at once, and each PEA can store two weight sub-tiles in its WBUF.
When the DTP is enabled, each PEA seamlessly performs two-tiled processing for an activation sub-tile, maximizing the reuse of the activation sub-tile and increasing DWO's utilization.
Under the DTP, the outer products of $\textbf{W}_{\text{LO}}\textbf{x}_{\text{LO}}$ for the second weight sub-tile can be allocated to DWOs to avoid the bounded throughput by few SWOs, as depicted in the first and second graphs of Fig. \ref{fig:design_option_eval}(b).
The DTP allows each PEA to produce two $4 \times 4$ partial-sums at a time, thus each PEA needs two CSs, two S-ACCs, and the doubled local partial-sum buffer.
Through the adoption of DTP, it is possible to achieve high utilization for both DWOs and SWOs, even at a high sparsity for weight HO slices.
In addition, it significantly reduces data movement from external memory to $\textit{Panacea}$ and from AMEM to the AQS-GEMM core by increasing the data reuse of activation sub-tiles, improving the energy efficiency and throughput of $\textit{Panacea}$ without additional outer-product calculators.

\begin{figure*}[t]
    \centering
    \includegraphics{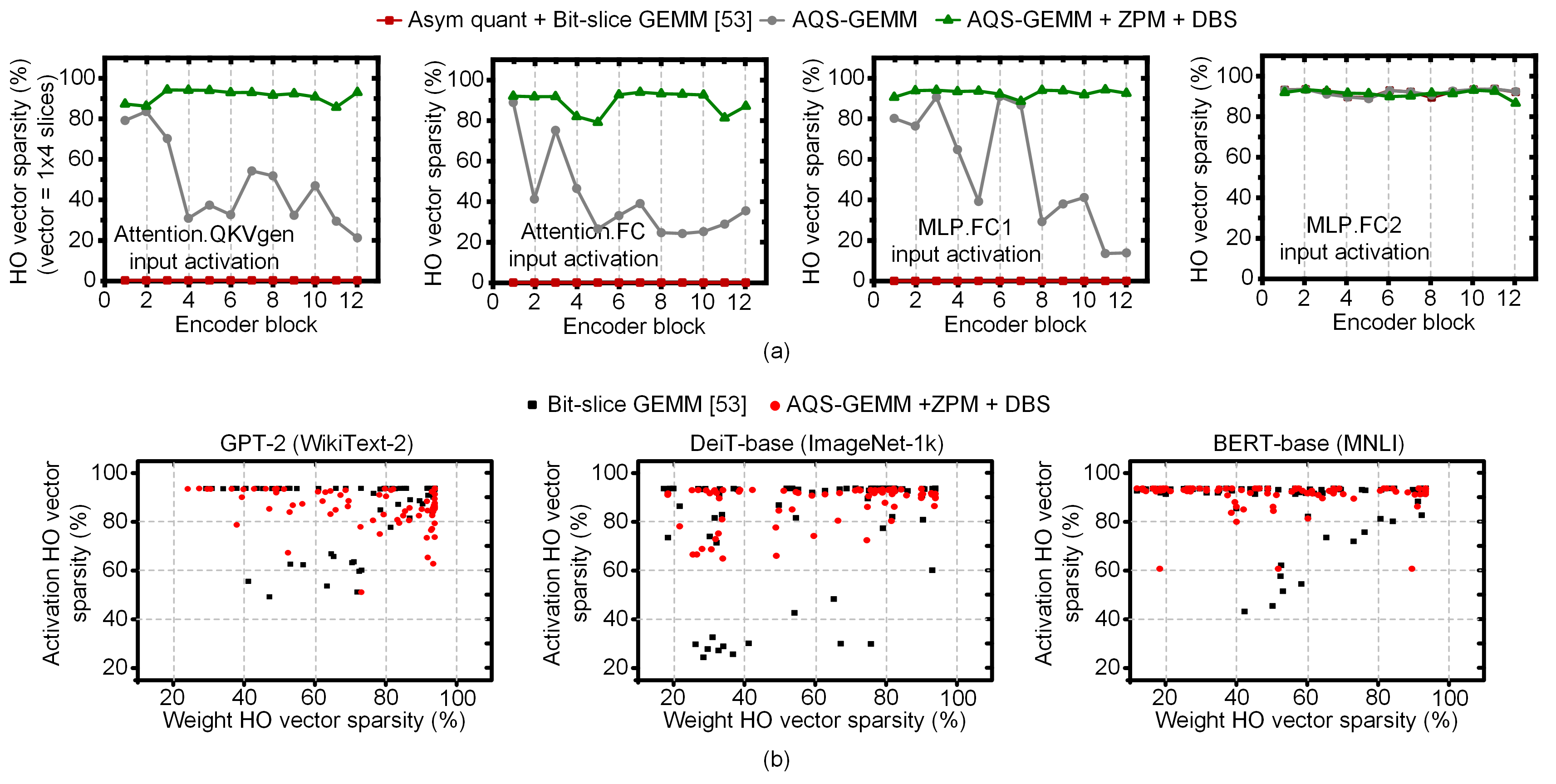}
    \vspace*{-3mm}
    \caption{(a) HO vector sparsity of activations in the DeiT-base model's layers (b) The vector sparsity of weight/activation HO slices observed by \textit{Sibia} \cite{im2024sibia} and \textit{Panacea} over all layers of the DNN benchmarks.}
    \label{fig:sparsity_eval}
    \vspace*{-4mm}
\end{figure*}

\section{Evaluation}
Our evaluations demonstrate \textit{Panacea}'s hardware efficiency, comparing it to the previous dense DNN accelerators including systolic array using weight stationary (SA-WS) and output stationary (SA-OS) \cite{asgari2020meissa}, SIMD \cite{keller202395}, and the recent bit-slice accelerator \textit{Sibia} \cite{im2024sibia}.
Note that SIMD, SA-WS, and SA-OS compute dense GEMM with 8-bit quantized weights and activations, while \textit{Panacea} and \textit{Sibia} use 4b$\times$4b multipliers to compute bit-slice GEMMs.
For a fair comparison, we make all designs to utilize identical design parameters: the number of multipliers, the size of on-chip SRAM, and the bandwidth between DRAM and accelerators.
More precisely, we assume that they utilize 3072 4b$\times$4b multipliers, 256-bit/cycle DRAM bandwidth, 192KB on-chip SRAM to store weights, input activations, and outputs, where an 8b$\times$8b multiplier is equivalent to four 4b$\times$4b multipliers.
To estimate the energy cost of external memory, we use the DRAM emulator, CACTI 7.0 \cite{balasubramonian2017cacti} and estimate the performance of designs based on a 28nm CMOS technology.
Given a specific accelerator architecture and its data-flow, we count the number of cycles and the number of activated modules during inference based on Hugging-Face open-source DNN models, considering bit-slice sparsity in real benchmarks. 
Then, we use the obtained data to estimate energy with the post-layout results of building blocks like multipliers, adders, and buffers.

\begin{figure*}[t]
    \centering
    \includegraphics{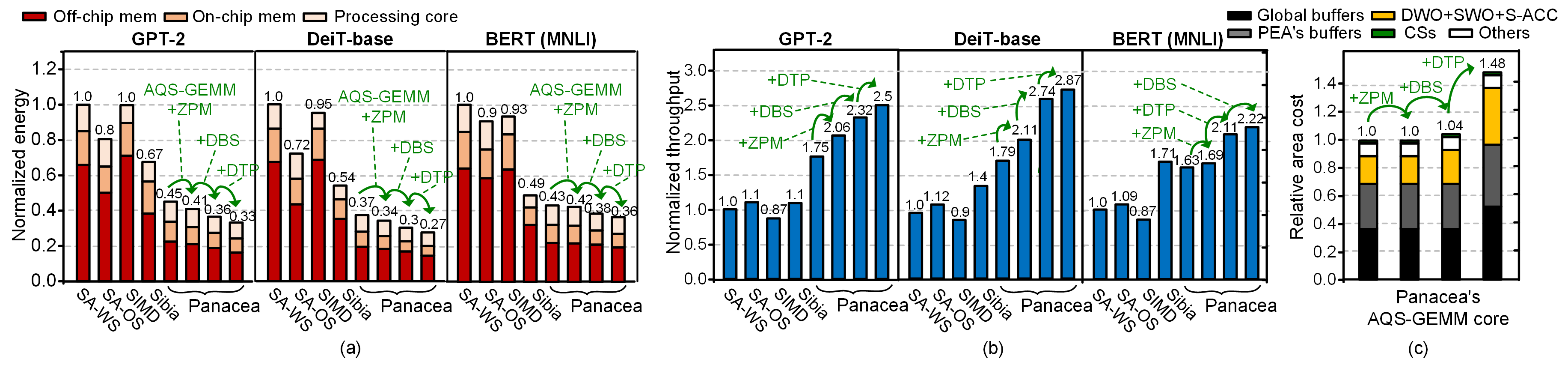}
    \vspace*{-5.5mm}
    \caption{(a) Energy breakdown and (b) throughput of the DNN accelerators for different benchmarks. (c) Relative area cost for applying the proposed methods. Note that we assume that all designs use identical hardware resources, such as the size of on-chip memory and the number of multipliers.}
    \label{fig:breakdown}
    \vspace*{-1mm}
\end{figure*}

\begin{figure*}[t]
    \centering
    \includegraphics{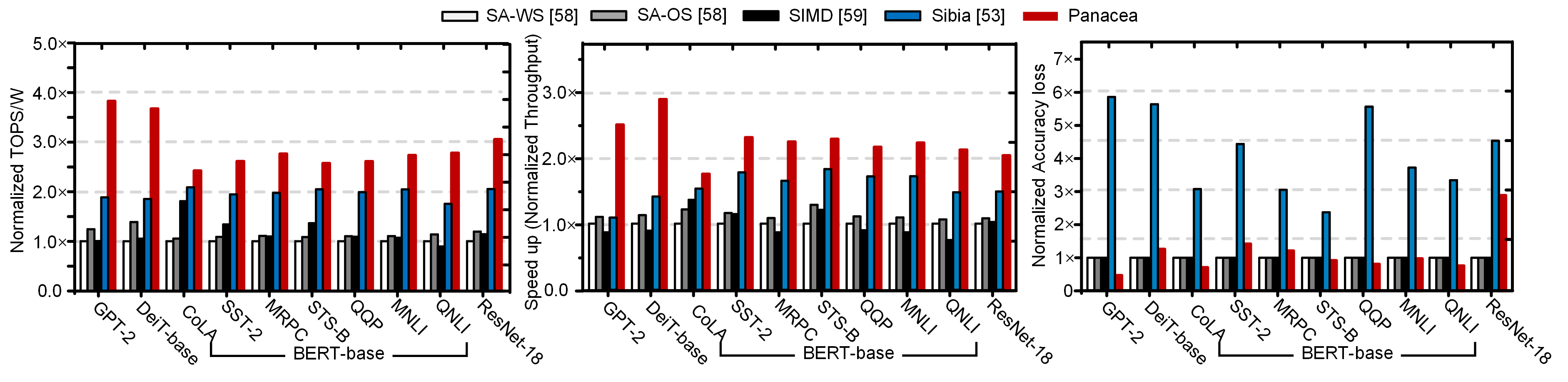}
    \vspace*{-5mm}
    \caption{Energy efficiency, throughput, and accuracy loss evaluations of \textit{Panacea} and previous accelerators \cite{asgari2020meissa, keller202395, im2024sibia} on different DNN models. Note that we assume that all designs use 192KB on-chip memory and 3072 4b$\times$4b multipliers, where an 8b$\times$8b multiplier is identical to four 4b$\times$4b multipliers.}
    \label{fig:evaluation}
\end{figure*}

\noindent\textbf{Throughput evaluation on different design options.}
Fig. \ref{fig:design_option_eval} shows \textit{Panacea}'s throughput improvement for different HO vector sparsities of weight/activation, compared to the previous accelerators \cite{asgari2020meissa, keller202395}. 
Fig. \ref{fig:design_option_eval}(a) and (b) present scenarios with 16 PEAs, each having different numbers of DWOs and SWOs. 
The graphs also illustrate the effect of DTP and different weight/activation sizes. 
More precisely, Fig. \ref{fig:design_option_eval}(a) shows throughput in the case of using 4 DWOs and 8 SWOs per PEA.
In situations where the HO slice sparsity of weight and activation is low, \textit{Panacea} achieves lower throughput than the previous designs \cite{asgari2020meissa, keller202395}.
This is because $\textbf{W}_{\text{HO}}\textbf{x}_{\text{HO}}$,$\textbf{W}_{\text{LO}}\textbf{x}_{\text{HO}}$,$\textbf{W}_{\text{HO}}\textbf{x}_{\text{LO}}$ become dense bit-slice GEMMs, and they are allocated to few DWOs, resulting in a long delay from DWOs.
Conversely, at the high slice sparsity, \textit{Panacea} achieves up to 3.7$\times$, 3.35$\times$, and 3.14$\times$ speedup when compared to SA-WS, SA-OS \cite{asgari2020meissa}, and SIMD \cite{keller202395}, respectively.
By applying the proposed DTP at high sparsity scenarios, \textit{Panacea}'s throughput is further improved by 1.11$\times$ than the disabled DTP mode.
Fig. \ref{fig:design_option_eval}(b) shows the throughput of \textit{Panacea} using 8 DWOs and 4 SWOs in each PEA.
Although the throughput is still lower than SIMD \cite{keller202395} design at very low sparsity, the use of more DWOs has narrowed the gap in dense conditions.
When the vector sparsity increases, the limited number of SWOs bounds the speedup.
In such cases, \textit{Panacea} provides higher throughput by adopting the proposed DTP.
However, as the size of weight increases, as illustrated in Fig. \ref{fig:design_option_eval}(b), the DTP starts to be enabled at higher vector sparsity due to the difficulty of meeting the DTP enable condition.
Compared to previous DNN accelerators \cite{verhelst2017embedded, asgari2020meissa, keller202395}, \textit{Panacea} achieves better speedup with large-sized weights and activations, significantly reducing off-/on-chip memory accesses due to its data-compression.
In the case of a small-sized activation, as shown in Fig. \ref{fig:design_option_eval}, all DNN accelerators can store the entire activation at once.
This easily avoids repeated data access and consequently reduces the effectiveness of data movement based on compression.
Ultimately, an appropriate number of DWOs and SWOs can be determined by analyzing slice-vector sparsity in actual DNNs, as depicted in Fig. \ref{fig:sparsity_eval}.

\noindent\textbf{Vector-level sparsity evaluation on DNN benchmarks.}
Since the AQS-GEMM applies compression at the vector level, as shown in Fig. \ref{proposed_main}, it is crucial to examine the sparsity at the vector level.
Fig. \ref{fig:sparsity_eval}(a) shows the HO vector sparsity of asymmetrically quantized activations in different GEMM methods, especially for DeiT-base\cite{touvron2021training}.
The previous bit-slice GEMMs \cite{im2024sibia, shomron2020non} cannot utilize any vector sparsity for most of the layers due to asymmetric quantization, which produces nonzero slices that cannot be skipped.
Note that a non-linear function in MLP produces many near-zero values that result in a lot of zero HO slices even with asymmetric quantization, so the previous bit-slice GEMM only achieves high vector sparsity in the activation of MLP.FC2 layer, as shown in Fig. \ref{fig:sparsity_eval}(a).
In contrast, the AQS-GEMM enables the vector sparsity for all layers by compressing frequent HO slices at the vector level.
The ZPM and DBS further increase the vector sparsity for wide distributions.

Fig. \ref{fig:sparsity_eval}(b) presents comparisons between the AQS-GEMM of \textit{Panacea} and the bit-slice GEMM of \textit{Sibia}\cite{im2024sibia} in terms of the weight/activation HO vector sparsity across all layers within three transformer models: DeiT-base\cite{touvron2021training}, BERT-base\cite{devlin2018bert}, and GPT-2\cite{radford2019language}\footnote{Bit-slice GEMMs adopt 10-bit symmetric weight quantization in MLP layers of GPT-2 \cite{radford2019language} to avoid accuracy loss, segmenting a 10-bit integer into three slices based on SBR.}.
Two methods utilize the identical SBR for the weight's bit-slicing, thus they achieve the same HO vector sparsity for weights.
While the bit-slice GEMM \cite{im2024sibia} achieves high HO vector sparsity in activations with symmetric quantization, the AQS-GEMM also achieves comparable vector sparsity with asymmetric quantization.
In several layers, the AQS-GEMM outperforms the bit-slice GEMM \cite{im2024sibia} by increasing HO vector sparsity through the ZPM and DBS.

\noindent
\textbf{Performance breakdown and trade-off analysis.}
To evaluate large-scale DNN benchmarks, \textit{Panacea} is designed with 4 DWOs and 8 SWOs in each PEA, as shown in Fig. \ref{fig:design_option_eval}, providing consistently high throughput across various levels of vector sparsity in weights.
This configuration is based on observations from Fig. \ref{fig:sparsity_eval}(b), where most activations have very high vector sparsity while weights have varying vector sparsity. 
Furthermore, given the large size of weights and activations in transformer models, using fewer DWOs successfully improves performance. 
Fig. \ref{fig:breakdown} shows the energy breakdown and the trade-off between energy, throughput, and area-overhead.
By handling only uncompressed slices within the AQS-GEMM, \textit{Panacea} significantly reduces the number of off/on-chip memory accesses and improves hardware-efficiency, compared to previous designs \cite{im2024sibia, asgari2020meissa, keller202395}.
\textit{Panacea}'s performance is further enhanced by gradually adopting the proposed the ZPM, DBS, and DTP.
For GPT-2 \cite{radford2019language} and the WikiText-2 \cite{merity2016pointer}, the ZPM further improves energy consumption by 10\% and throughput by 17\%, increasing slice-level sparsity without any area overhead.
The DBS also improves energy consumption by 11\% and throughput by 12\%, requiring small area overhead due to its additional shifting units in the AQS-GEMM core.
By enhancing hardware utilization for high slice-vector sparsity in DNN layers, the DTP reduces energy consumption by 8.9\% and improves throughput by 7.6\%.
However, it requires additional buffers and larger on-chip memory, resulting in some area overhead.

\noindent\textbf{Accelerator performance for DNN models.}
Fig. \ref{fig:evaluation} presents results for the various large transformer-based models \cite{devlin2018bert, touvron2021training, radford2019language} and a small non-transformer-based model \cite{he2016deep}.
In specific, for GPT-2\cite{radford2019language} and the WikiText-2 dataset \cite{merity2016pointer} including large-sized activations, \textit{Panacea} achieves 3.82$\times$, 3.07$\times$, 3.81$\times$, and 2.03$\times$ higher energy efficiency (TOPS/W) than SA-WS, SA-OS \cite{verhelst2017embedded, asgari2020meissa}, SIMD \cite{keller202395}, and \textit{Sibia} \cite{im2024sibia}, respectively, successfully achieving high activation sparsity based on the proposed ZPM and DBS.
For BERT-base\cite{devlin2018bert} and the GLUE\cite{wang2018glue} dataset, which uses fewer tokens, \textit{Panacea} still outperforms the dense DNN accelerators \cite{asgari2020meissa, keller202395} by achieving 2.67$\times$ and 2.41$\times$ higher energy efficiency in average, respectively, due to the compressed data movement from external memory to the AQS-GEMM core.

\begin{figure}[t]
    \centering
    \vspace*{-4mm}
    \includegraphics{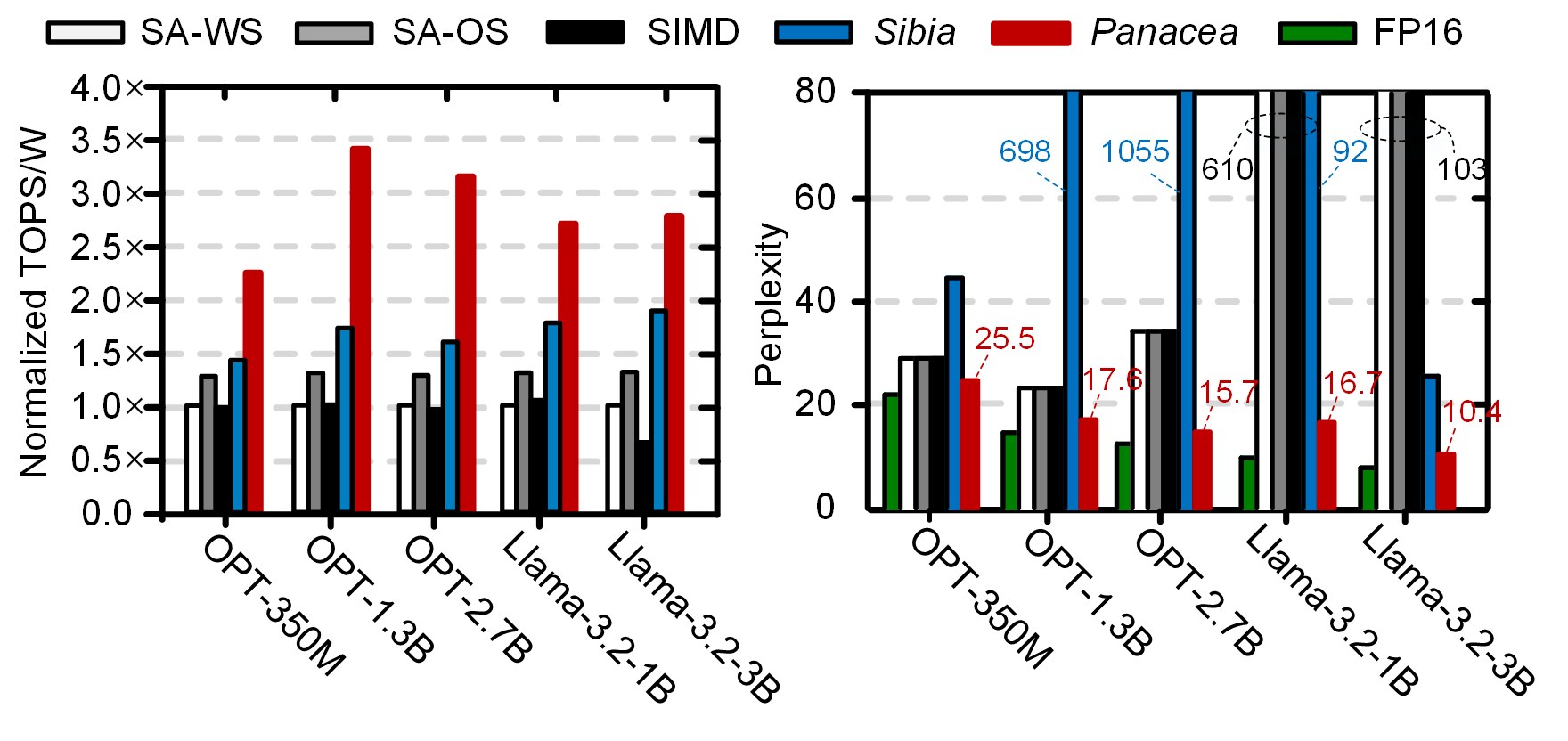}
    \vspace*{-2.0mm}
    \caption{Energy efficiency and perplexity evaluations of \textit{Panacea} and previous DNN accelerators \cite{asgari2020meissa, keller202395, im2024sibia} on different LLMs.}
    \label{fig:LLM}
\end{figure}

\begin{figure}[t]
    \centering
    \includegraphics{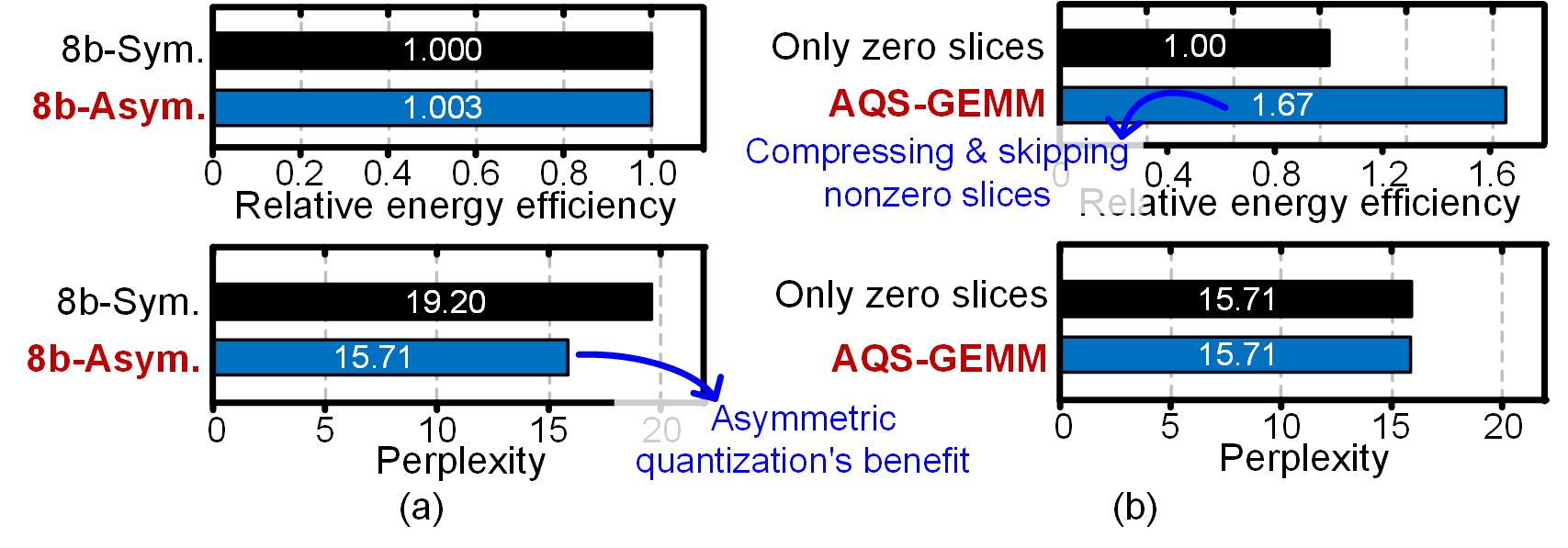}
    \vspace*{-1.5mm}
    \caption{Relative energy efficiency and perplexity for OPT-2.7B\cite{zhang2022opt} when (a) using different quantization methods in \textit{Panacea} and (b) using the same asymmetric quantization in two versions of \textit{Panacea}.}
    \label{fig:decouple}
    \vspace*{-2.0mm}
\end{figure}

\begin{figure}[t]
    \centering
    \vspace*{-1mm}
    \includegraphics{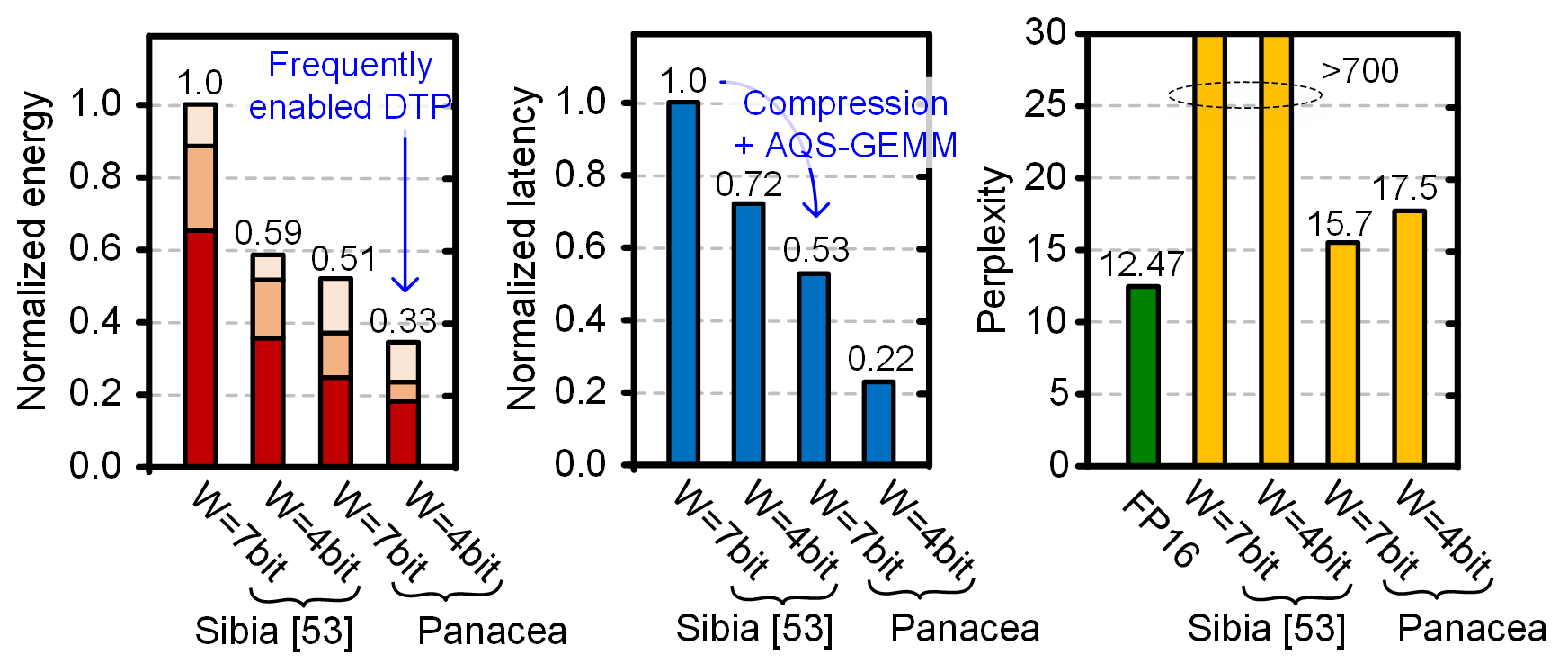}
    \vspace*{-1.0mm}
    \caption{Energy breakdown, latency, and perplexity of \textit{Sibia} \cite{im2024sibia} and \textit{Panacea} when they use 4-bit or 7-bit weights for OPT-2.7B \cite{frantar2022optq, zhang2022opt}.}
    \label{fig:low-bit}
\end{figure}

\begin{figure}[t]
    \centering
    \vspace*{-0.5mm}
    \includegraphics{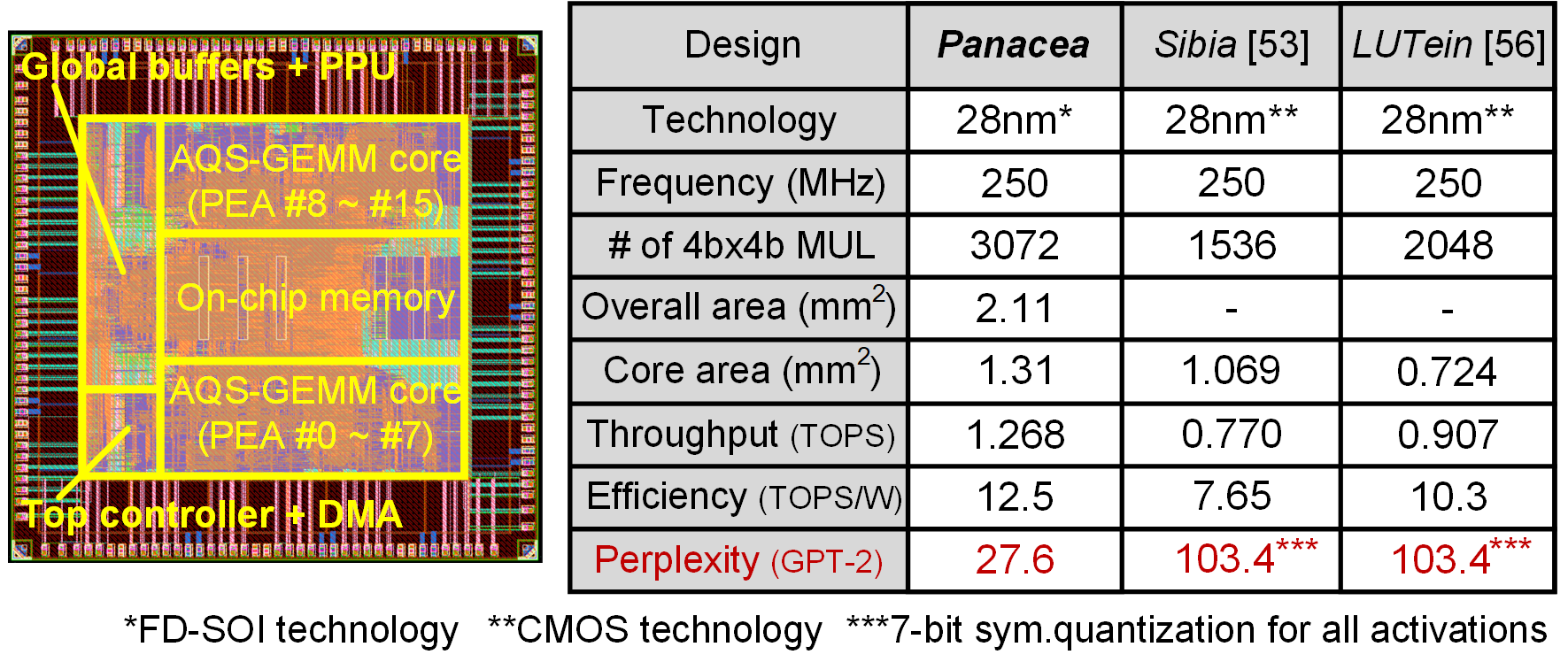}
    \vspace*{-5.0mm}
    \caption{The layout of $\textit{Panacea}$ and ASIC-level comparison table.}
    \label{fig:ASIC}
    \vspace*{-1.5mm}
\end{figure}

Fig. \ref{fig:LLM} presents the evaluation results for various large language models (LLMs)\cite{zhang2022opt, llama3.2} using the WikiText-2 dataset\cite{merity2016pointer}.
\textit{Panacea} achieves 1.57$\times$, 1.97$\times$, and 1.96$\times$ higher energy efficiency compared to \textit{Sibia} in the OPT models \cite{zhang2022opt} with 350M, 1.3B, and 2.7B parameters, respectively, while maintaining perplexity (PPL) similar to FP16.
Unlike the OPT models\cite{zhang2022opt}, the Llama-3.2 models\cite{llama3.2} are more challenging to quantize weights without PPL loss due to structural differences and large outliers.
To address these challenges, we applied OPTQ \cite{frantar2022optq}, a weight-only quantization, and 64 channel-wise quantization to weights.
Especially in the case of \textit{Sibia} \cite{im2024sibia} and \textit{Panacea}, which support mixed-precision quantization, the inputs to sensitivity-critical layers, i.e., the down-projection layer, can be expressed with three bit-slices, resulting in better PPL than dense GEMM architectures\cite{ asgari2020meissa,keller202395}.
For the Llama-3.2 3B model, \textit{Panacea} achieves 2.77$\times$, 2.11$\times$, 4.24$\times$, and 1.47$\times$ greater energy efficiency compared to the SA-WS, the SA-OS \cite{asgari2020meissa}, the SIMD \cite{keller202395}, and \textit{Sibia} \cite{im2024sibia}, respectively, even when considering mixed-precision quantization.

\noindent
\textbf{Decoupling the advantages of asymmetric quantization and the AQS-GEMM.}
\textit{Panacea} supports not only asymmetric quantization for activations, but also symmetric quantization by setting every zero-point to 128 within the 8-bit range. 
Fig. \ref{fig:decouple}(a) compares the two quantization methods on \textit{Panacea} for OPT-2.7B\cite{zhang2022opt} and shows that while asymmetric quantization yields a lower PPL, the energy efficiency and the throughput remain nearly equivalent due to the ZPM and DBS maximizing the slice sparsity. 
Additionally, Fig. \ref{fig:decouple}(b) demonstrates that \textit{Panacea}, utilizing the AQS-GEMM to skip both nonzero and zero slices, improves energy efficiency by $1.67\times$ and throughput by $2.10\times$ compared to skipping only zero slices, while both achieve the same PPL by producing exact results in the AQS-GEMM.


\noindent
\textbf{Accelerator performance in low-bit quantization.}
As mentioned in Section III-B, \textit{Panacea} divides a $(3n+4)$-bit weight into $(n+1)$ 4-bit slices and a $(4k+4)$-bit activation into $(k+1)$ 4-bit slices. 
Note that \textit{Sibia}\cite{im2024sibia} uses symmetrically quantized activations, limiting its representation to $(3k+4)$ bits, despite using $(k+1)$ 4-bit slices.
Fig. \ref{fig:low-bit} compares the performance of \textit{Sibia} \cite{im2024sibia} and \textit{Panacea}, on OPT-2.7B for $n=0$ or $n=1$ with $k=1$. 
The 4-bit quantization for weights ($n=0$) was handled with OPTQ\cite{frantar2022optq}, which prevents PPL loss in extreme-low-bit quantization ($\leq 4$-bit).
\textit{Panacea} consumes only 56\% of energy compared to \textit{Sibia}\cite{im2024sibia} as the DTP is frequently enabled due to the 4-bit weights without HO slices.
\textit{Panacea} further outperforms \textit{Sibia}\cite{im2024sibia} by achieving 1.9$\times$ and 3.3$\times$ lower latency for 7-bit and 4-bit weights, respectively, while still maintaining the acceptable PPL.

\noindent
\textbf{ASIC implementation.}
\textit{Panacea} is implemented in a 28nm FD-SOI technology.
Fig. \ref{fig:ASIC} shows its layout and a comparison table summarizing implementation results of the recent bit-slice DNN accelerators: \textit{Sibia}\cite{im2024sibia}, \textit{LUTein}\cite{im2024lutein}, and \textit{Panacea}.
To support the proposed methods and $2\times$ more multipliers, \textit{Panacea} requires a small overhead in terms of the core area.
Due to supporting the proposed AQS-GEMM and sparsity optimization methods, \textit{Panacea} simultaneously achieves better throughput, energy efficiency, and algorithm-level performance on large-scale DNN models.


\section{Conclusion}
This paper proposes a bit-slice DNN accelerator, \textit{Panacea}, that enables the bit-slice GEMM with asymmetric quantization.
\textit{Panacea}'s AQS-GEMM compresses high-order slices frequently observed in asymmetric quantization and skips their computations.
To improve the efficiency of AQS-GEMM, \textit{Panacea} presents two algorithm-hardware co-optimization methods: ZPM and DBS, which increase slice-vector sparsity for activations.
To efficiently support the proposed methods, each processing element of \textit{Panacea} dedicates two types of operators and utilizes the specialized dataflow, maximizing data reuse.
Consequently, \textit{Panacea} provides attractive hardware/algorithm performances during DNN inferences.
Our DNN evaluation code will be open-sourced at https://github.com/Dongyunkam/Panacea.

\section{Acknowledgments}
We express our gratitude to the anonymous reviewers for their valuable comments and insightful suggestions.
This work was supported in part by Korea Institute for Advancement of Technology (KIAT) grant funded by the Korea Government (MOTIE) (P0017304, Human Resource Development Program for Industrial Innovation), in part by the National Research Foundation of Korea (NRF) grant funded by the Korea government (MSIT) (No. 2022R1A2C2092521), in part by the NAVER-Intel Co-Lab, and in part by the IC Design Education Center (IDEC).


\bibliographystyle{unsrt}
\bibliography{main}

\end{document}